\documentclass[preprint,12pt,authoryear]{elsarticle}

\usepackage{amssymb}
\usepackage{amsmath}

\usepackage{overpic}
\usepackage{xcolor}
\usepackage{physics}
\usepackage{siunitx}
\usepackage{wasysym}
\usepackage{ulem}

\newcommand\abeqn[2]{\refstepcounter{equation}
\[\label{#1}
#2
\eqno{(\theequation{\mathrm{a,b}})}
\]}

\newcommand\abceqn[2]{\refstepcounter{equation}
\[\label{#1}
#2
\eqno{(\theequation{\mathrm{a,b,c}})}
\]}

\newcommand\xyeqn[2]{
\[
\label{#1}
#2
\]}

\DeclareMathOperator{\arctanh}{arctanh}

\journal{International Journal of Solids and Structures}

\begin{document}

\begin{frontmatter}

\title{Bridging a gap: A heavy elastica between point supports} 

\author[label1]{Grace~K.~Curtis}
\author[label1]{Ian~M.~Griffiths} 
\author[label1]{Dominic~Vella\corref{cor1}}
\ead{dominic.vella@maths.ox.ac.uk}
\cortext[cor1]{Corresponding author}

\affiliation[label1]{organization={Mathematical Institute, University of Oxford},
            addressline={Woodstock Rd}, 
            city={Oxford},
            postcode={OX2 6GG}, 
            country={United Kingdom}}

\begin{abstract}
 We study the deformation and slip-through of a heavy elastic beam suspended above two point supports and subject to an increasing body force --- an idealized model of a fibre trapped in the pores of a filter as flow strength increases, for example. Using both asymptotic and numerical techniques, we investigate the behaviour of the beam under increasing body force and the maximum force that can be supported before it must slip between the supports. We quantify this maximum body force as a function of the separation between the two supports. Surprisingly, we show the existence of a critical separation below which the beam can withstand an arbitrarily large body force, even in the absence of friction. This is understood as the limit of a catenary between the supports that is connected to (and supported by the tension in) a vertically hanging portion outside the supports. We explore how frictional forces impact the deformation and load-bearing capacity of the beam and show that our results are consistent with laboratory experiments.
\end{abstract}




\end{frontmatter}

\section{Introduction}

Flexible objects in contact with a rigid boundary and subject to a body force occur in a wide variety of scenarios, whether in an industrial setting or in day-to-day life: the main cable in a suspension bridge \citep{Zhang2017bridgefric,Zhou2019bridgecat,Zhang2021bridgevib}, buckling of a pipe within a borehole when drilling for oil \citep{Miller2015buckling,Miller2015lockup}, and the rucking of a rug \citep{vella2009ruck,Kolinski2009}, are some examples.
While many of these examples are at quite large scales, the essential ingredients (a uniform body force and a flexible slender structure) can be a useful approximation of complex scenarios at small scales. Two examples of the same ingredients at smaller scales are the flow-induced trapping of a fibre in a filter \citep{lant2022filter}, and a soft robot walking on the floor \citep{shepherd2011,hu2018magelastic,wu2023caterpillar}.

When modelling the deformation of elastic materials, it is important to maintain key features such as the flexibility, as well as the forces underpinning the deformation. In all of the aforementioned examples, the deformation is influenced by contact with a surface. Therefore it is important to understand the effects of the frictional and restoring forces as well as a constant body force on the material.
However, a general contact region can be complex and so it is especially convenient that several of the earlier examples, such as the trapping of fibres and soft robot locomotion, can be approximated as point contact. 
 
When considering a supported elastic beam, the question of when the beam will slip through the supports and fall is often of interest. 
For example, when considering a model for a point supported or variable-arc-length beam, the question is whether an equilibrium configuration exists \citep{Athisakul2008val, Chen2010supports, Plaut2011supports}. The aim of this class of problem is to understand the critical load that can be withstood by the beam before it slides off the support, and how this depends on the location of the support.

Previous work on point-supported elastic beams has focused on investigating the optimal positions with which to lift an elastic cable \citep{Wang1990lifting}, as well as the critical weight and support span with which to store flexible pipes or cables \citep{Chen2010supports}. The two models differ from one another as the first model concerns the lifting of a cable, equivalent to having only a nonzero vertical reaction force, whereas the latter captures how a cable is supported, taking into account a nonzero horizontal component of the reaction force. 
In either case, the focus has often been on determining either the critical support positioning for a given load or,  for a given support positioning, the critical load  at which the beam slides off. 

This system also has important practical implications. For example, when considering the filtration of fibres that have detached from the clothes in a washing machine or tumble-dryer, the slipping of a fibre through the filter mesh occurs due to the action of the fluid flow through the filter. Such a process is undesirable as this corresponds to failure in the filter operation. 

Motivated by the filtration of fibres, in this paper, we consider the deformation of a supported elastica under an increasing body force, starting from a small-deformation state. We note that, for highly-deformed fibres, the force due to an external flow will actually vary with arc length along the fibre, since the force experienced by a fibre is proportional to the length perpendicular to the external flow. However, for model simplicity we will assume a uniform body force along the length. This is sufficient to uncover the qualitative behaviour of the system; generalizing to an arc-length-dependent force is straightforward. We also note that the model describes the deformation of an elastic beam that is loaded uniformly across its length with changes in load only happening quasi-statically.

In Section~2, we present our model of the deformation of a beam under increasing force, for the case in which the contact between the beam and the supports is assumed to be frictionless. We study the behaviour and load-bearing capacity of the beam, above which the beam must slip and fall between the supports. We show that for  sufficiently close supports (supports a fraction $<1/e$ of the total length of the beam apart) the system is able to support an arbitrary high load in equilibrium --- since $1/e\approx0.37$, this seems to reproduce a result of \citet{Chen2010supports} but allows additional insight that may be generalized to include friction.

In Section~3, we generalize the model to explore how frictional forces impact the deformation and load-bearing capacity of the beam. 
One complication that this introduces to the problem is that friction allows a body to hold a range of configurations \citep{vella2009ruck,Plaut2011flat,Plaut2011supports}.  In Section~4 we present results from experiments while in Section~5 we draw conclusions on our investigation of this point-supported heavy beam.

\section{A heavy beam on point supports}

\subsection{Model setup}

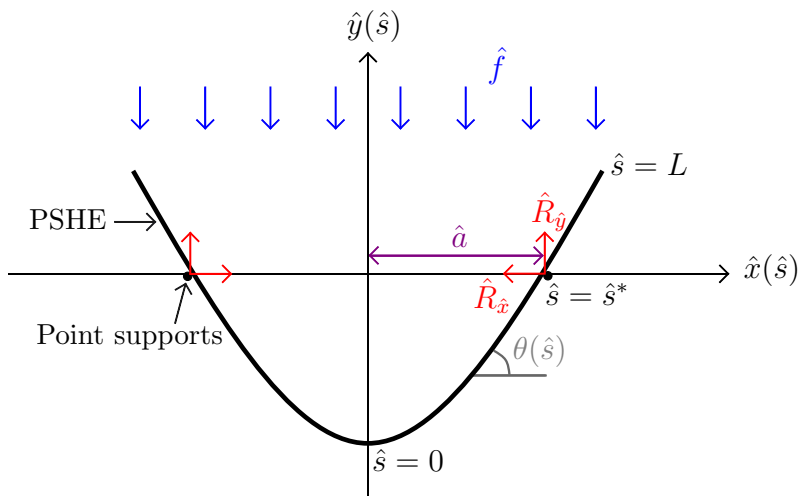
\begin{figure}[ht]
\centering
\vspace{4mm}
\begin{overpic}[width=.7\textwidth,tics=10]{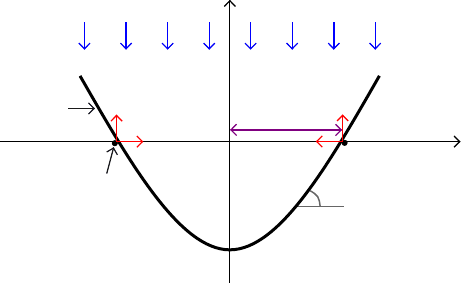}
\put(47,64){$\hat{y}(\hat{s})$}
\put(102,30.2){$\hat{x}(\hat{s})$}
\put(50.5,3.5){$\hat{s}=0$}
\put(74.5,26.5){$\hat{s}=\hat{s}^*$}
\put(83.5,44.5){$\hat{s}=L$}
\put(70,19){\textcolor{gray}{$\theta(\hat{s})$}}
\put(66.5,58){\textcolor{blue}{$\hat{f}$}}
\put(61.5,34.5){\textcolor{violet}{$\hat{a}$}}
\put(72.5,38){\textcolor{red}{$\hat{R}_{\hat{y}}$}}
\put(64.5,26){\textcolor{red}{$\hat{R}_{\hat{x}}$}}
\put(4,21){\small{Point supports}}
\put(3,37){\small{PSHE}}
\end{overpic}
\vspace{2mm}
\caption{Sketch of the setup for the point-supported heavy elastica (PSHE) of length $2L$ under the action of a body force $\hat{f}$ per unit length and supported by points located a distance $2\hat{a}$ apart.}\label{fig:setup_sketch}
\end{figure}

We consider a beam of length $2L$, placed symmetrically on two point supports, which are set a horizontal distance $2\hat{a}$ apart, as shown in figure~\ref{fig:setup_sketch}. Here we will assume that the contact between the beam and the supports is frictionless; we revisit this assumption in Section~3. The deformation occurs under the action of a body force $\hat{f}$ per unit length, which is intended to represent the weight of the beam; the system is therefore referred to as the Point Supported Heavy Elastica (or PSHE) henceforth. However, it may also represent the action of an idealized viscous force from a uniform flow perpendicular to the beam. 

The coordinates $(\hat{x},\hat{y})$ are defined such that the point supports lie on $\hat{y}=0$, and the centre of the PSHE lies at $\hat{x}=0$. We define~$\theta$ as the angle the tangent to the PSHE makes with the positive $\hat{x}$-axis. The arc length $\hat{s}$ is defined symmetrically such that $\hat{s}=0$ at the centre of the PSHE, $\hat{s}=\pm\hat{s}^*$ at the two point supports, and $\hat{s}=\pm L$ at the two free ends. The point supports provide a reaction force with vertical and horizontal components, $\hat{R}_{\hat{y}}$ and $\hat{R}_{\hat{x}}$, respectively, defined in the directions as shown in figure~\ref{fig:setup_sketch}. The signs here are chosen such that when the PSHE is in the shape given by figure~\ref{fig:setup_sketch}, both reaction forces are positive.

Geometry relates the cartesian coordinates $(\hat{x},\hat{y})$ to the intrinsic coordinates $(\hat{s},\theta)$ via 
\abeqn{eqn2:full_dim_system_xy}{\dv{\hat{x}}{\hat{s}} = \cos\theta, \hspace{30mm} \dv{\hat{y}}{\hat{s}} = \sin\theta.}
We exploit the symmetry of the problem to consider only the half of the PSHE with $0 \leq \hat{s} \leq L$. A vertical force balance then gives
\begin{align}
    \dv{\hat{s}}(\hat{t} \sin\theta + \hat{n}\cos\theta) &= \hat{f} - \hat{R}_{\hat{y}} \delta(\hat{s}-\hat{s}^*), \label{eqn2:full_dim_system_vert}
\end{align} 
where $\hat{t}$ and $\hat{n}$ are the tension and shear forces, defined to be tangent and normal to the PSHE, respectively, and $\delta(\cdot)$ is the Dirac $\delta$-function. A horizontal force balance gives
\begin{align}
    \dv{\hat{s}}(\hat{t} \cos\theta - \hat{n}\sin\theta) &= \hat{R}_{\hat{x}} \delta(\hat{s}-\hat{s}^*).\label{eqn2:full_dim_system_horz}
\end{align} 
Finally, a moment balance gives
\begin{align}
    EI \dv[2]{\theta}{\hat{s}} + \hat{n} &= 0,\label{eqn2:full_dim_system_mom}
\end{align}
where $E$ is the Young's modulus of the PSHE and $I$ is the second moment of inertia of its cross-section \cite[see][for example]{Howell2008solid}.

These equations are to be solved subject to the following boundary conditions. From symmetry, we require that
\abeqn{eqn2:dim_BCs_symm}{\theta(0)=0, \hspace{40mm} \hat{x}(0) = 0.}
 \noindent
At the contact point, we must have the following continuity conditions:
\abeqn{eqn2:dim_BCs_cont}{\hat{x}(\hat{s}^*) =\hat{a}, \hspace{39mm} \hat{y}(\hat{s}^*)=0,}
\vspace{-7mm}
\xyeqn{}{[\hat{x}]_{\hat{s}=\hat{s}^*} = 0, \hspace{37mm} \left[\hat{y} \right]_{\hat{s}=\hat{s}^*} = 0, \hspace{2mm} \eqno{(\theequation{\mathrm{c,d}})}}
\vspace{-4mm}
\xyeqn{}{[\theta]_{\hat{s}=\hat{s}^*} = 0, \hspace{32.5mm} \left[\dv{\theta}{\hat{s}} \right]_{\hat{s}=\hat{s}^*} = 0, \hspace{2mm} \eqno{(\theequation{\mathrm{e,f}})}}
where $[\cdot]$ denotes the jump across $\hat{s}=\hat{s}^*$.
Finally, at the free boundary we have 
\abeqn{eqn2:dim_BCs_free}{\hspace{1.5mm} \hat{t}(L)=0, \hspace{40mm}  \hat{n}(L)=0,}
\vspace{-5mm}
\xyeqn{}{\dv{\theta}{\hat{s}} {(L)}=0. \hspace{60mm}\eqno{(\theequation{\mathrm{c}})}}

We can immediately remove the $\delta$-functions from equations (\ref{eqn2:full_dim_system_vert}) and (\ref{eqn2:full_dim_system_horz}) by integrating equations (\ref{eqn2:full_dim_system_vert}) and (\ref{eqn2:full_dim_system_horz}) over the point support. This gives jump conditions for $\hat{t}$ and $\hat{n}$ at the support, effectively splitting our system over two regions: central ($0 \leq \hat{s} \leq \hat{s}^*$) and overhang ($\hat{s}^* \leq \hat{s} \leq L$).
Integrating (\ref{eqn2:full_dim_system_vert}) over $0 \leq \hat{s} \leq L$, using (\ref{eqn2:dim_BCs_symm}a), (\ref{eqn2:dim_BCs_free}a,b), and $\hat{n}(0)=0$ (by symmetry), we find $\hat{R}_{\hat{y}}=\hat{f}L$. 
This represents a global vertical force balance: $\hat{f}L$ is the total force on the (half) PSHE, which is balanced by the vertical component of the reaction force from one support.
Since the contact between the PSHE and the point supports is frictionless, we impose that $\hat{t}$ is continuous over the point support, i.e.~$\left[\hat{t}\right]_{\hat{s}=\hat{s}^*}=0$. Therefore, integrating (\ref{eqn2:full_dim_system_vert}) over $\hat{s}^{*^-} \leq \hat{s} \leq \hat{s}^{*^+}$, we find  $\left[\hat{n}\right]_{\hat{s}=\hat{s}^*}=-\hat{f} L \sec\theta(\hat{s}^*)$, with $\hat{R}_{\hat{x}}=\hat{f} L \tan\theta(\hat{s}^*)$ from (\ref{eqn2:full_dim_system_horz}).

Taking the half-length $L$ to be the natural length scale, we nondimensionalize by letting,
\begin{align}
    \hat{s}&=L s, && \hat{x}=L x, && \hspace{1.3cm}\hat{y}=L y, \\(\hat{n},\hat{t})&=\frac{EI}{L^2}(n,t), && \hat{f}=\frac{EI}{L^3}f, && (\hat{R}_{\hat{x}}, \hat{R}_{\hat{y}}) = \frac{EI}{L^2} (R_x,R_y). \label{eqn2:non-dim scaling}
\end{align}
This non-dimensionalization introduces an important dimensionless parameter (in addition to the force $f$): the half-gap  $a=\hat{a}/L$. 

Rewriting, the system of dimensionless equations to be solved in \linebreak $0\leq s<s^*$ and $s^*<s\leq 1$ is 
\begin{subequations}
\begin{align}
    \dv{x}{s} &= \cos\theta, \\
    \dv{y}{s} &= \sin\theta, \label{eqn2:dimensionless_y} \\
    \dv{s}(t \sin\theta + n\cos\theta) &= f, \label{eqn1: y_force_bal}\\
    \dv{s}(t \cos\theta - n\sin\theta) &= 0,\label{eqn1: x_force_bal}\\
    \dv[2]{\theta}{s} + n &= 0, 
\end{align} \label{eqn2:dimensionless_eqns}%
\end{subequations}
with boundary conditions 
\abeqn{eqn2:dimensionless_BCs}{\theta(0)=0, \hspace{40.5mm} x(0) = 0, \hspace{-1.25mm}}
\vspace{-6.5mm}
\xyeqn{}{x(s^*) =a, \hspace{39mm} y(s^*)=0, \hspace{0.5mm} \eqno{(\theequation{\mathrm{c,d}})}}
\vspace{-5mm}
\xyeqn{}{[x]_{s=s^*} = 0, \hspace{37mm} \left[y \right]_{s=s^*} = 0, \hspace{2.5mm} \eqno{(\theequation{\mathrm{e,f}})}}
\vspace{-3mm}
\xyeqn{}{[\theta]_{s=s^*} = 0, \hspace{32mm} \left[\dv{\theta}{s}\right]_{s=s^*} = 0, \hspace{2.5mm} \eqno{(\theequation{\mathrm{g,h}})}}
\vspace{-2.5mm}
\xyeqn{}{ [t]_{s=s^*}=0, \hspace{24.5mm} [n]_{s=s^*}=-f \sec\theta^*, \eqno{(\theequation{\mathrm{i,j}})}}
\vspace{-5mm}
\xyeqn{}{\hspace{1mm} t(1)=0, \hspace{40mm} n(1)=0, \eqno{(\theequation{\mathrm{k,l}})}}
\vspace{-5mm}
\xyeqn{}{\dv{\theta}{s} {(1)}=0, \hspace{60mm} \eqno{(\theequation{\mathrm{m}})}}
where we have denoted $\theta(s^*)$ by $\theta^*$. Note that (\ref{eqn2:dimensionless_eqns}) is a twelfth-order system with the unknown $s^*$, making it thirteenth-order; (\ref{eqn2:dimensionless_BCs}) provides the thirteen boundary conditions required. 

\subsection{Numerical approach}
\label{sec2:numerics}

To solve the two-point boundary value problem (\ref{eqn2:dimensionless_eqns}) with unknown contact arc length $s^*$, we scale both the $0 \leq s \leq s^*$ and $s^* \leq s \leq 1$ regions onto $[0,1]$ by introducing two new variables, 
\begin{align}
    \eta=\frac{s}{s^*}, && \xi=\frac{s-s^*}{1-s^*}. \label{eqn:eta_tau_rescale}
\end{align}  
The price of this conversion is that the relevant equations differ by different scaling factors. To solve the full system, we therefore need to consider separate solutions in the central and overhang regions for each variable.

\subsubsection{Force versus displacement control}
There are several ways in which we can solve the problem using the boundary value problem solver \texttt{bvp4c} in \textit{MATLAB}, depending on whether we imagine the force or displacement to be controlled.

\paragraph{Force control}
The first, and arguably more intuitive, method is to impose the force $f$ and half-gap $a$, and solve for the unknown contact arc length $s^*$. In this case, we start by solving for small $f$, and then increase $f$ via continuation. 
\paragraph{Displacement control}
The second method is to impose the contact arc length $s^*$ and half-gap $a$, and solve for the unknown force $f$ required to give this solution. In this case, we start from $s^*$ close to $a$, solve for $f$, and then increase $s^*>a$ up to $s^*=1$ via continuation.

It is also possible to fix $f$ and then decrease or increase $a$ values, with unknown $s^*$, again via continuation, but we do not consider this here.  

Since we are generally interested in the solutions under increasing force, our primary interest is in the force-controlled approach. However, we shall make use of both methods to trace the system's evolution. 
An initial guess for each of the twelve variables and the remaining unknown parameter (either $s^*$ or $f$, depending on the chosen control parameter) is required by the solver. Focusing on increasing force $f$, we assume the PSHE starts from a small-deformation state; this allows us to use a small-angle approximation to provide an analytic guess for the numerical solver.

\subsection{Small-angle approximation}
\label{sec2:small_angle}

We begin by assuming that the deformation angle is small, which we know will occur for $f \ll 1$.  Since the deformation of the PSHE from the $x$-axis is small, we must also have that $y$ is small. Hence, $x \approx s$ and, in particular, the contact arc length is approximately the same as the half-gap, i.e.~$s^* \approx a$. We linearize the governing equations and hence expect the other variables to scale with $f$. Hence, we let 
\begin{align}
    \theta=f \tilde{\theta},  \qquad n=f\tilde{n}, \qquad y=f \tilde{y}.
\end{align}
Expanding the system (\ref{eqn2:dimensionless_eqns}) and boundary conditions (\ref{eqn2:dimensionless_BCs}) to leading order in $f$ and solving the resulting (linear) system, we find the solution (expressed in the original dimensionless variables) to be
\begin{subequations}
\begin{align}
        s^*&\approx a, \\ x&\approx s, \\ t(x)&\approx 0, \\
    n(x)&\approx f
    \begin{cases}
        x & 0\leq x<s^*, \\
        x-1 & s^*<x\leq 1,
    \end{cases} \\
    \theta(x) &\approx \frac{f}{6}
    \begin{cases}
        -x^3+3(2s^*-1)x & 0\leq x<s^*, \\
        -x^3+3x^2-3x+3{s^*}^2 & s^*<x\leq 1, \label{eqn2:small_angle_theta}
    \end{cases}\\
    y(x) &\approx  \frac{f}{24}
    \begin{cases}
        -x^4+6(2s^*-1)x^2+({s^*}^4-12{s^*}^3+6{s^*}^2) & 0\leq x<s^*, \\
        -x^4+4x^3-6x^2+12{s^*}^2 x +({s^*}^4-16{s^*}^3+6{s^*}^2) & s^*<x\leq 1.
    \end{cases}  \label{eqn2:small_angle_shape}
\end{align} \label{eqn2:small_angle_solutions}%
\end{subequations} 
\vspace{-4mm}

Using this asymptotic result, we can also find an approximation for the contact arc length $s^*$ as follows
\begin{align*}
    s^*=\int^a_0 \left[ 1+(y'(x))^2 \right]^{1/2} ~\dd x \approx a + \frac{1}{2}\int^a_0 (y'(x))^2 ~\dd x.
\end{align*}
Computing this using (\ref{eqn2:small_angle_shape}) for $x~<~s^*$, we find that 
\begin{align}
    s^* \approx a + f^2 \left[ \frac{1}{504}a^7 - \frac{1}{30}a^6 + \frac{11}{60}a^5 - \frac{1}{6}a^4 + \frac{1}{24}a^3 \right]. \label{eqn2:small_angle_arclength}
\end{align} 

While the asymptotic solution (\ref{eqn2:small_angle_solutions})-(\ref{eqn2:small_angle_arclength}) provides some intuitive insight into the shape (e.g.~the curvature of the central region seems to change sign values when $s^*\approx1/2$, i.e.~around $a\approx1/2$), it is important as a starting point for the solution of the full system (\ref{eqn2:dimensionless_eqns}), allowing us to evolve from a small-deformation state where $f \ll 1$ and $s^* \approx a$.

\subsection{Results}

\subsubsection{Shapes of the PSHE}

\begin{figure}[ht]
\centering
\begin{overpic}[width=.995\textwidth,tics=10]{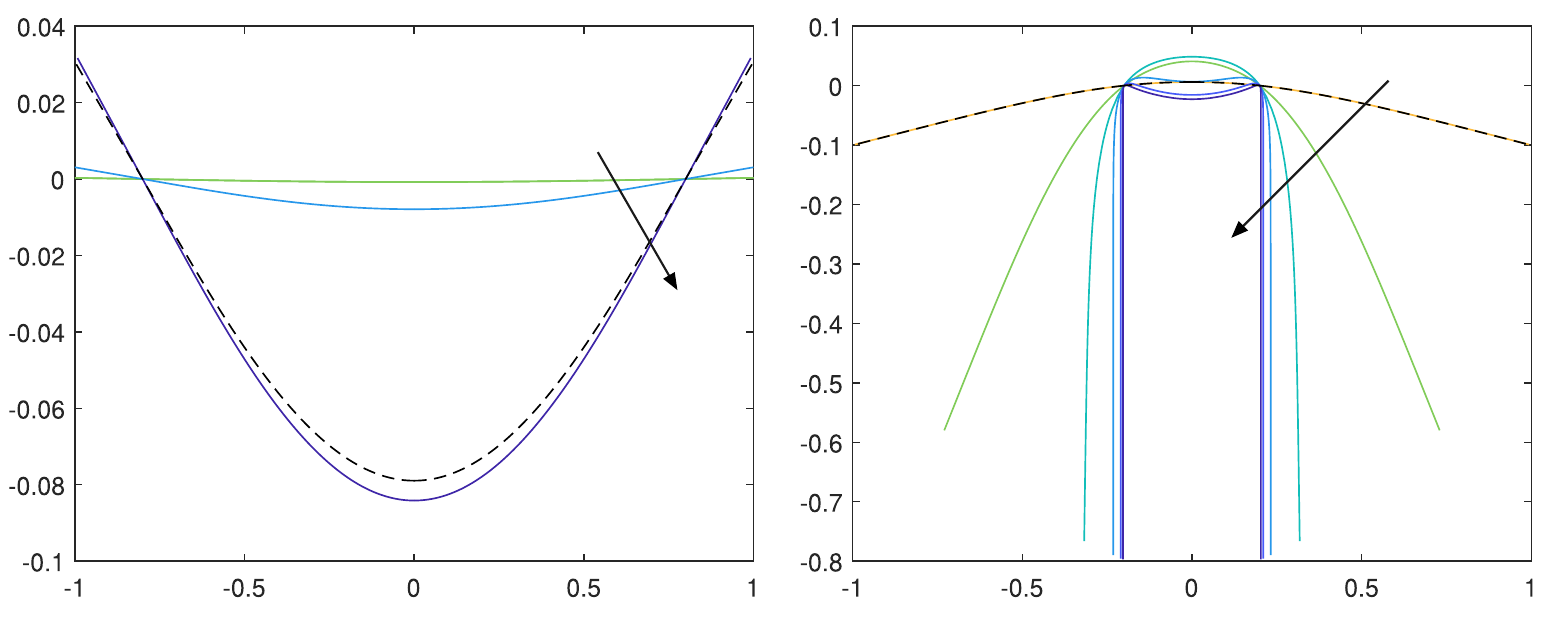}
\put(-6,21){\small{$y(s)$}}
\put(24.25,-0.5){\small{$x(s)$}}
\put(75,-0.5){\small{$x(s)$}}
\put(27,33){\small{Increasing $f$}}
\put(82.5,36.5){\small{Increasing $f$}}
\put(25,-5){(a)}
\put(75.5,-5){(b)}
\end{overpic}
\vspace{2mm}
\caption{Plots of the PSHE shape for various values of the half-gap, $a$, and increasing force, $f$. The dashed curves represent the small-deformation shape, $y(x)/f$, or (equivalently) the shape predicted with $f=1$ by the small-deformation theory. Solid curves show the numerically-computed shapes for: (a) $a=0.8$, $\log_{10}f=\{-2,-1,0\}$. (b) $a=0.2$, $\log_{10}f~=~\{0,1,2,3,4,5\}$. The different shapes illustrated are referred to as `$u$'-shape (as in (a)) or `$m$' and `$n$'-shape (as in (b)).}
\label{fig:shape_snapshots}
\end{figure}
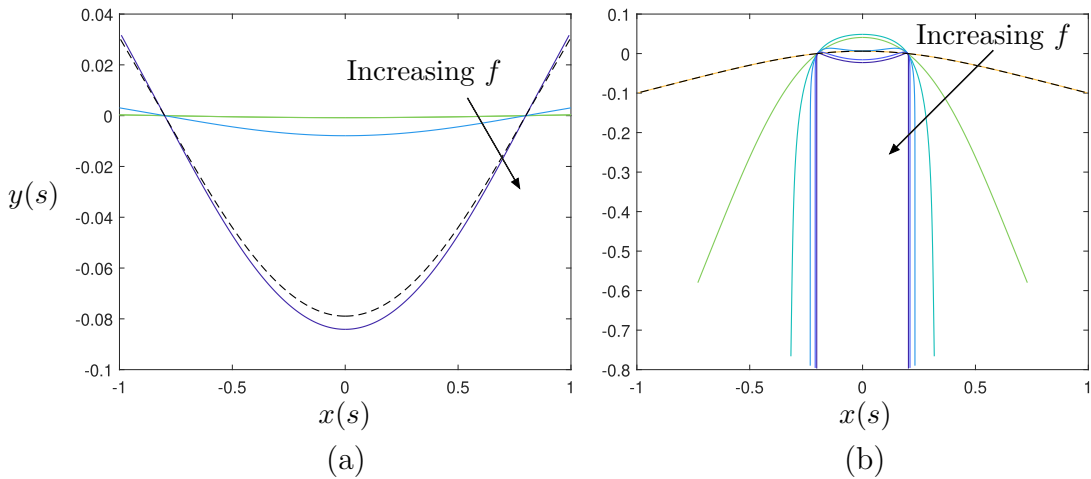

Typical shapes of the PSHE as the force $f$ is increased for various values of half-gap $a$, are shown in figure~\ref{fig:shape_snapshots}. As expected, the numerically-determined shapes diverge from the small-angle shapes as $f$ increases beyond unity. 

An interesting feature shown in figure~\ref{fig:shape_snapshots}  is that the shape of the PSHE passes through three possible phases depending on the values of both $f$ and $a$. 
Following the similarity of the shapes shown in figure~\ref{fig:shape_snapshots} to the respective letters, we refer to each shape as a `$u$', `$m$', or `$n$'-shape. For $a=0.8$, for example, we find that the PSHE seems only to realize a $u$-shape, while for $a=0.2$, the PSHE seems to realize both an $n$-shape and $m$-shape under increasing force, but not a $u$-shape (figure~\ref{fig:shape_snapshots}).  We define each shape concretely by considering the following two features: the sign of the curvature at the centre of the PSHE, and the sign of the angle at the endpoint of the PSHE. In particular, we define:
\begin{align*}
    \text{(i) }& u\text{-shape: }  && \theta'(0)>0, && \theta(1)>0. \qquad \\
    \text{(ii) }& m\text{-shape: }  && \theta'(0)>0, && \theta(1) \leq 0. \qquad \\
    \text{(iii) }& n\text{-shape: }  && \theta'(0) \leq 0, && \theta(1) \leq 0. \qquad
\end{align*}

 Figure~\ref{fig:shape_snapshots} also highlights other features of the problem. For example, with $a=0.8$, as we continue to increase the value of $f$, numerical solutions cease to exist for sufficiently large $f$. This suggests the existence of a maximum force, beyond which the PSHE cannot sustain an equilibrium shape. In practice, we identify this with the PSHE slipping through the supports. We also notice that the PSHE can support a much greater force when $a=0.2$ than it can when $a=0.8$, which suggests that this maximum force must also be dependent on $a$. In the next section, we explore the existence and value of this maximum force, $f_\mathrm{max}(a)$, in more detail.

\subsubsection{Non-monotonicity of $f$}

\begin{figure}[ht]
\centering
\begin{overpic}[width=.7\textwidth,tics=10]{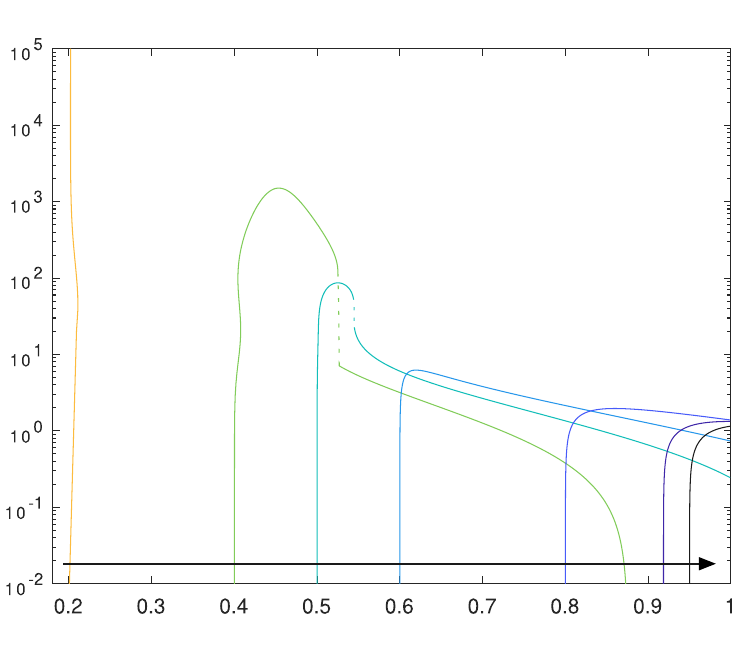}
\put(50,1){\small{$s^*$}}
\put(-4,38){\small{$f$}}
\put(12.5,14){\small{Increasing $a$}}
\end{overpic}
\caption{Plots of the force, $f$, against the contact arc length, $s^*$, for half-gaps $a~=~\{0.2,0.4,0.5,0.6,0.8,0.9185,0.95\}$.}
\label{fig:f_vs_xi_combined}
\end{figure}

Naively, one might expect that the force, $f=f_\mathrm{max}$, needed to push the PSHE between the two supports would correspond to the value of $f$ such that the entire length of the PSHE lies between the supports (i.e.~$s^*=1$). Treating $s^*$ as the control parameter and determining the corresponding value of $f$, we see that $f(s^*;a)$ is generally non-monotonic (figure~\ref{fig:f_vs_xi_combined}): for $f<f_{\mathrm{max}}$, two possible solutions may exist, one with $s^*<s^*_{\mathrm{crit}}$ and one with $s^*>s^*_{\mathrm{crit}}$, where $s^*_{\mathrm{crit}}$ is defined by $f(s^*_{\mathrm{crit}})=f_\mathrm{max}$.
This non-monotonicity separates our solutions into two branches; our interest in the deformation of a PSHE under increasing force means we will focus on the branch with $s^*<s^*_{\mathrm{crit}}$ since this branch can be accessed from the small $f$ asymptotic solution. This differs from the approach of \cite{Chen2010supports} who considered both solution branches.

We find that $s^*_{\mathrm{crit}}=1$ is only realized when the half-gap $a \gtrsim 0.9185$, as seen in figure~\ref{fig:f_vs_xi_combined} --- for these values of $a$, $f(s^*;a)$ is monotonic. The shapes realized through this deformation are similar to that of the $u$-shapes in figure~\ref{fig:shape_snapshots}(a), except that the endpoints of the PSHE eventually reach the supports, at which point $f=f_{\mathrm{max}}$.

For $a=0.4$, for example, we see the shape transition from an $n$-shape to a $m$-shape under increasing $f$, as seen in figure~\ref{fig:shape_snapshots}(b) for $a=0.2$.
This requires a large increase in force, and the contact arc length decreases as the transition occurs, as shown in (figure~\ref{fig:f_vs_xi_combined}).
In such cases, we find the curve in figure~\ref{fig:f_vs_xi_combined} by first using force control to increase $f$ until the PSHE no longer realizes an $n$-shape, and then switching to displacement control, increasing $s^*$. A similar transition is realized in the $s^*>s^*_{\mathrm{crit}}$ branch for $a=0.4$, where the shape snaps from an $m$-shape to a $u$-shape as $f$ decreases. This is seen numerically as a jump in $f$, however, since we are interested in the deformation of the PSHE under increasing force, we do not consider this further.

We can also see from figure~\ref{fig:f_vs_xi_combined} that the value of $f_\mathrm{max}$ increases as the half-gap $a$ decreases. More surprisingly, however, for $a=0.2$ we note that it appears there is no finite $f_\mathrm{max}$; there seems to be a critical half-gap $a_\infty \in [0.2,0.4)$, below which there does not exist a finite maximum force.

\subsubsection{Behaviour for large $f$ and $a\leq a_\infty$}
\label{sec2:large_f}

To understand the behaviour of the PSHE in the limit of large forces, $f \gg 1$, (and to understand if there is, indeed, no maximum load $f_{\mathrm{max}}$ in some configurations), it is natural to rescale the forces by $f$, i.e.~we let
\begin{align}
    n = fN, && t=fT.
\end{align}
With this rescaling, the force and moment balance equations (\ref{eqn2:dimensionless_eqns}c--e) become:
\begin{subequations}
\begin{align}
    \dv{s}(T \sin\theta + N\cos\theta) &= 1 , \label{eqn2:large_f_vert}\\
    \dv{s}(T \cos\theta - N\sin\theta) &= 0, \label{eqn2:large_f_horz}\\
    \frac{1}{f} \dv[2]{\theta}{s} + N &= 0, \label{eqn2:large_f_mom}
\end{align} \label{eqn2:large_f_system}%
\end{subequations} 
subject to the boundary conditions (\ref{eqn2:dimensionless_BCs}), where (\ref{eqn2:dimensionless_BCs}i--l) are rescaled.
This system is now almost independent of the loading $f$, though we note that in (\ref{eqn2:large_f_mom}) there is a small parameter, $1/f$, in front of the highest derivative. As such, we expect a boundary layer-type solution.
In this case, from equation (\ref{eqn2:large_f_mom}), we have $N=0$ everywhere except in the boundary layer and hence expect to have a heavy string, i.e.~the catenary. This is similar to what we see in the numerical solution for large $f$.
We begin by considering the outer solution.

\paragraph{Outer solution: the full catenary}
\label{sec2:outer_sol}

For an outer solution, we will look at system (\ref{eqn2:large_f_system}) at leading order. Equation (\ref{eqn2:large_f_mom}) gives $N=0$ while (\ref{eqn2:large_f_vert}) and (\ref{eqn2:large_f_horz}) yield, respectively,
\abeqn{eqn2:outer_sol_system}{\dv{s}(T \sin\theta) = 1, \hspace{30mm} \dv{s}(T \cos\theta) = 0.} 
Equations (\ref{eqn2:outer_sol_system}) can be integrated once to give
\abeqn{eqn2:t_sin_cos_cat}{T \sin\theta =  s - \alpha, \hspace{35mm}
    T \cos\theta = \beta,}
where $\alpha$, $\beta$ are constants of integration. Here, $\alpha$ corresponds to the arc length position at which $\theta=0$, which in our notation is $s=0$; hence, $\alpha=0$. We discuss the physical significance of $\beta$ in due course. We note that there are two solutions for $T$ depending on whether we impose boundary conditions at $s=0$ or $s=1$. This means we must consider solutions in the central (i.e.~$0\leq s<s^*$) and overhang (i.e.~$s^*<s\leq 1$) regions separately, which we denote by subscripts $c$ and $o$, respectively. From equations (\ref{eqn2:t_sin_cos_cat}), and using  $y(s^*)=0$, we find that the shape and tension in the central region are: 
\begin{align}
    y_c &=\sqrt{s^2+\beta^2}-\sqrt{{s^*}^2+\beta^2}, \label{eqn2:outer_y_cat_sol} \\
    T_c &= \sqrt{s^2+\beta^2}, \label{eqn2:t_outer_c}
\end{align} 
where we note that $T(0)=\beta$.
This shape corresponds to the classic catenary, which may be seen by relating the arc length $s$ to $x$, i.e.
\begin{align}
    s&=\beta \sinh \left(\frac{x}{\beta}\right) \label{eqn2:arc_length_cat}
\end{align}
to yield
\begin{align}
    y_{c} &= \beta \left[\cosh \left(\frac{x}{\beta}\right) - \cosh \left(\frac{a}{\beta}\right) \right], \\
    T_{c} &= \beta\cosh \left(\frac{x}{\beta}\right).
\end{align}

For the solution in the overhang region, imposing $T(1)=0$ in equations (\ref{eqn2:t_sin_cos_cat}), gives $\theta_o(s)=-\pi/2$. This is a vertically hanging chain with solution 
\begin{align}
    y_{o} &= s^*-s, \label{eqn2:outer_y_o_sol}\\
    T_{o} &= 1-s, \label{eqn2:t_outer_o} 
\end{align}
where we have also used the boundary condition that $y(s^*)=0$. 

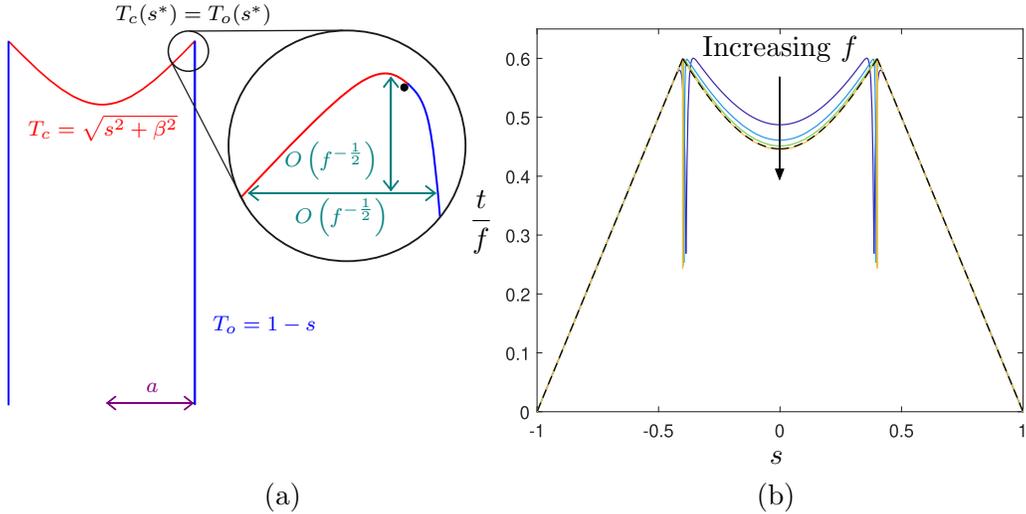
\begin{figure}[ht]
\centering
\begin{overpic}[width=.995\textwidth,tics=10]{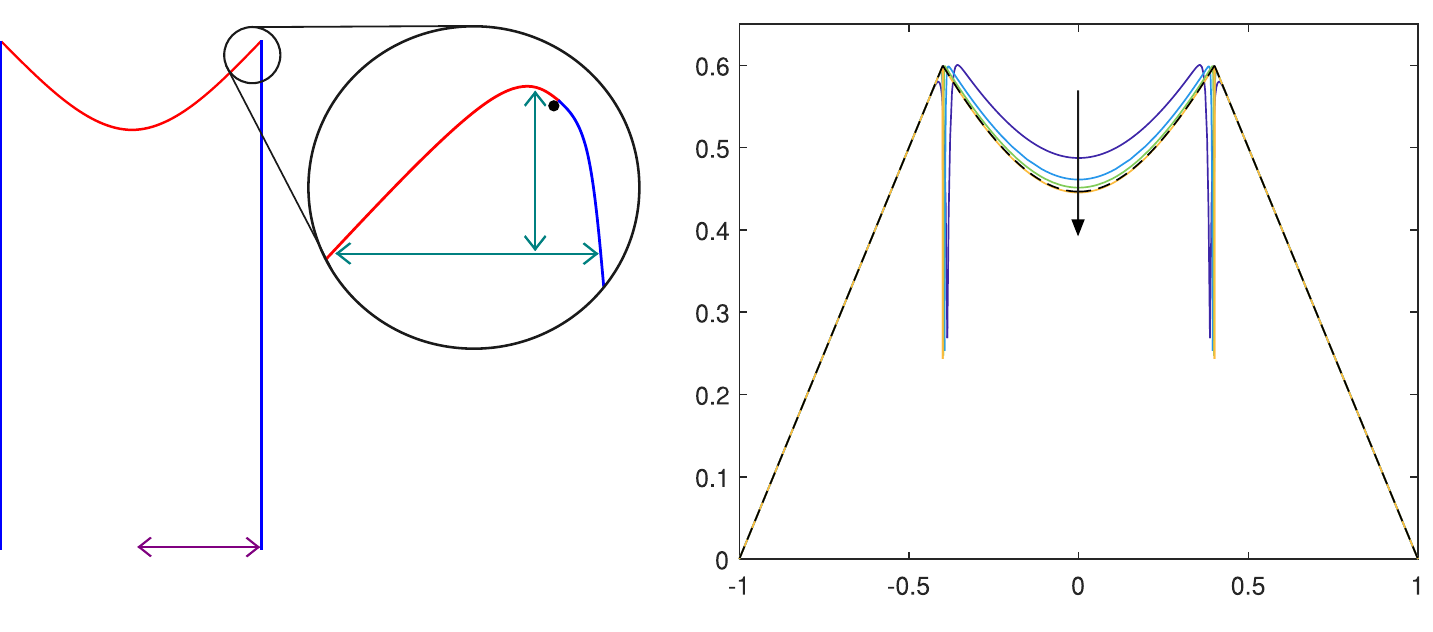}
\put(28,23.5){\scriptsize{\textcolor{teal}{$O\left(f^{-\frac{1}{2}}\right)$}}}
\put(27,29){\scriptsize{\textcolor{teal}{$O\left(f^{-\frac{1}{2}}\right)$}}}
\put(2,32){\scriptsize{\textcolor{red}{$T_c=\sqrt{s^2+\beta^2}$}}}
\put(20,13){\scriptsize{\textcolor{blue}{$T_o=1-s$}}}
\put(10.5,43.25){\scriptsize{$T_c(s^*)=T_o(s^*)$}}
\put(13.5,7){\scriptsize{\textcolor{violet}{$a$}}}
\put(45,23){\small{$\dfrac{t}{f}$}}
\put(74.25,0){\small{$s$}}
\put(67.75,39.75){\small{Increasing $f$}}
\put(25,-4){\small{(a)}}
\put(73,-4){\small{(b)}}
\end{overpic}
\vspace{1mm}
\caption{(a) Sketch of the shape of the PSHE near the supports in the limit $f \gg 1$. (b) Plots of the numerically determined tension, $t(s)/f$, as a function of arc length $s$ for half-gap $a~=~0.36$ (solid curves), for  $\log_{10}(f)~=~\{4, 5, 6, 7\}$, alongside $T(s)$ from the outer solution (\ref{eqn2:t_outer_c}) and (\ref{eqn2:t_outer_o}) (dashed curve).}\label{fig:outer_sol}
\end{figure}

We therefore have two solutions of the problem with $N=0$. These must match at $s=s^*$; in particular, imposing that the tension is continuous at the support, we have 
\begin{align}
    \sqrt{{s^*}^2+\beta^2} = 1 - s^*. 
    \label{eqn2:outer_sol_b_eqn}
\end{align}
However, this incurs a discontinuity in $\theta$ at the support that must be resolved in the boundary layer (figure~\ref{fig:outer_sol}(a)).

Before we consider the boundary layer at $s=s^*$, we first make some observations about the two solutions we have just found.

First, we note that the force $f$ does not enter into the solution (\ref{eqn2:outer_y_cat_sol})-(\ref{eqn2:t_outer_c}) and (\ref{eqn2:outer_y_o_sol})-(\ref{eqn2:t_outer_o}): when $N=0$, an arbitrarily large $f$ may be supported in this way. (We shall see shortly that it is only possible to find a suitable $\beta$ if $a<a_\infty=1/e$, however.)

Secondly, by combining equations (\ref{eqn2:outer_sol_system}), and using equation (\ref{eqn2:dimensionless_y}), since $T(1)=0$,  (\ref{eqn2:dimensionless_BCs}i), and $T$ and $y$ are both continuous over the supports, we can write $T(s)=y(s)-y(1)$. In particular
\begin{align}
    T(0)=y(0)-y(1), \label{eqn2:outer_sol_diff}
\end{align}
i.e.~the difference in height between the centre of the PSHE, and the end of the PSHE, is equal to the tension of the PSHE at the centre. Since we saw from eqn \eqref{eqn2:t_outer_c} that $\beta=T(0)$, eqn \eqref{eqn2:outer_sol_diff} gives us new insight into the classic catenary solution, (\ref{eqn2:outer_y_cat_sol}) and (\ref{eqn2:t_outer_c}); in particular, the physical meaning of the constant of integration $\beta$ appears not to be well known. Since combining \eqref{eqn2:outer_sol_diff} with \eqref{eqn2:t_outer_c} we have that $\beta = y(0)-y(1)$, one natural interpretation is that $\beta$ represents the distance between the lowest point of a catenary and the lowest point of the hanging chain needed to support that catenary. Since the hanging chain is usually omitted, this is not apparent: by adding the freely hanging part of the chain, the meaning of $\beta$ becomes clearer. For this reason we refer to the combination of hanging and central portions of the chain as the `full catenary'.

\paragraph{An analytic expression for $a_\infty$} 
We have seen that if a value of $\beta$ can be found such that equation (\ref{eqn2:outer_sol_b_eqn}) is satisfied, the PSHE can withstand an arbitrarily large force. We will determine the critical value of half-gap that separates this case from the case in which equilibria exist only when $f<f_\mathrm{max}$.

Substituting for $s^*$ from (\ref{eqn2:arc_length_cat}), we find that $\beta$ satisfies
\begin{align}
    a = -\beta \log \beta. \label{eqn2:a_inner_sol}
\end{align}
Crucially, this equation for $\beta$ has solutions provided that
\begin{align}
    a \leq \frac{1}{\mathrm{e}}:= a_\infty.
\end{align}
(The right-hand-side of (\ref{eqn2:a_inner_sol}) has a maximum at $\beta=1/\mathrm{e}$.)
Therefore, the PSHE can assume this catenary-type configuration, and hence withstand an infinite body force, provided that $a \leq a_\infty=1/\mathrm{e}$.  

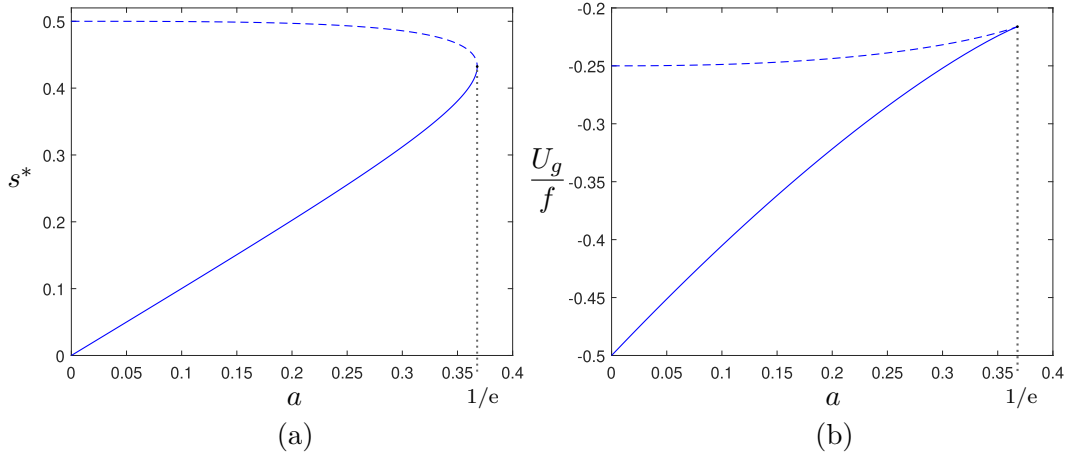
\begin{figure}[ht]
\centering
\begin{overpic}[width=.995\textwidth,tics=10]{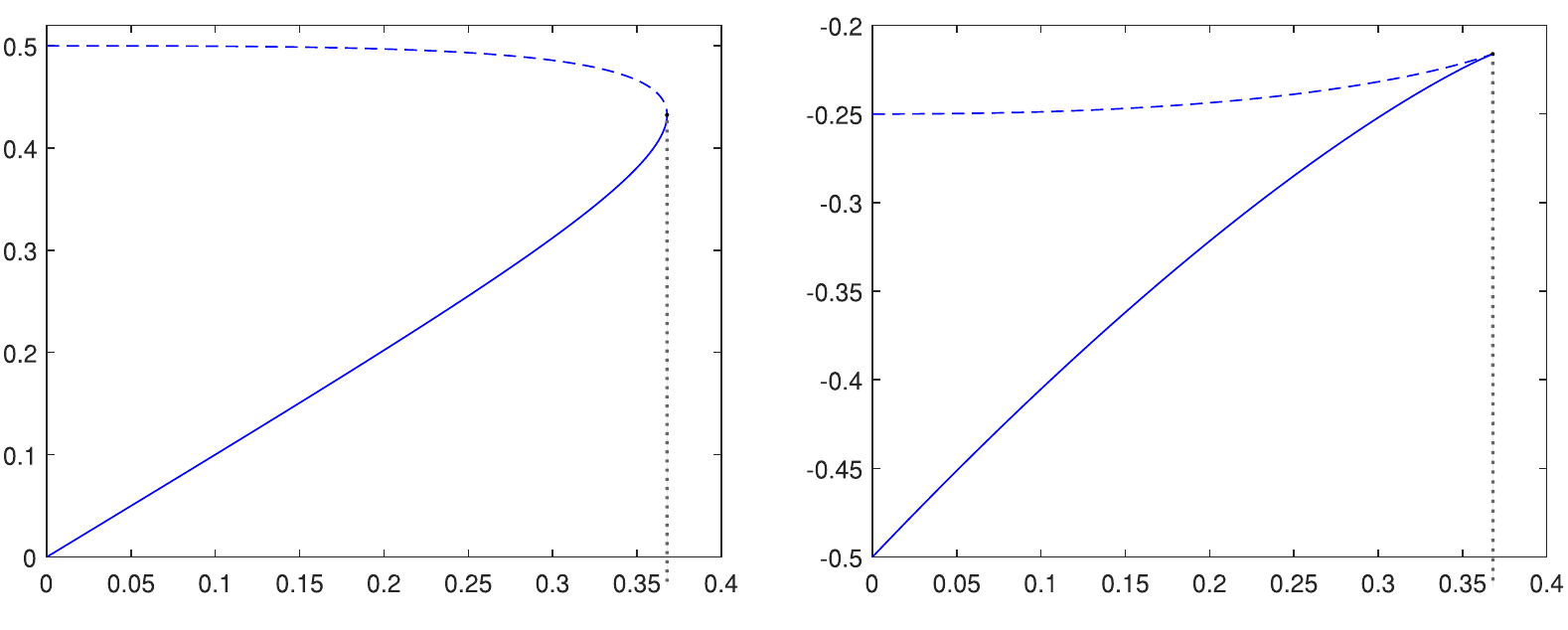}
\put(-3.25,21){\small{$s^*$}}
\put(24,-0.25){\small{$a$}}
\put(47.5,21){\small{$\dfrac{U_g}{f}$}}
\put(76.5,-0.25){\small{$a$}}
\put(23,-4){\small{(a)}}
\put(75.5,-4){\small{(b)}}
\put(41.5,0){\scriptsize{$1/\mathrm{e}$}}
\put(94,0){\scriptsize{$1/\mathrm{e}$}}
\end{overpic}
\vspace{1mm}
\caption{(a) Plot of the contact arc length, $s^*$, in the outer solution, (\ref{eqn2:arc_length_cat}), against half-gap, $a \leq 1/\mathrm{e}$, (\ref{eqn2:a_inner_sol}). (b) Plot of the (rescaled) potential energy, $U_g/f$, (\ref{eqn2:inner_sol_energy}), against half-gap, $a$, (\ref{eqn2:a_inner_sol}). The solid curves represent the branch of solutions with $s^*< s^*(a_{\infty})$, and dashed curves those with $s^*>s^*(a_{\infty})$. The limiting point $a=1/\mathrm{e}$ is shown by the vertical dotted line.}\label{fig:energy_outer}
\end{figure}

However, we note that the solution of (\ref{eqn2:a_inner_sol}) for $\beta$ is multi-valued for $a<a_{\infty}$, that is, for a given value of half-gap, there are two possible configurations that satisfy the equations: one with contact arc length $s^*< s^*(a_{\infty})$ and the other with $s^* > s^*(a_{\infty})$ (figure~\ref{fig:energy_outer}(a)). The existence of two solutions is unsurprising, as previously we realized only one solution due to our focus on solving the problem under increasing force, while the analysis above only holds for $f \gg 1$. To understand which solution to take, we will look at the energy of both solutions, and take the solution with less energy.

\paragraph{Energy and stability}
The total energy of the PSHE combines the potential and bending energies. In dimensional units, we have \linebreak $\hat{U}_\mathrm{total}=EI\left[ U_g + U_b \right]/L$, where
\begin{align}
    U_g = f\int^1_0 y(s) ~\dd s && \text{and} && U_b = \frac{1}{2} \int^1_0 \left(\dv{\theta}{s}\right)^2 ~\dd s,
\end{align}
are the dimensionless potential and bending energy, respectively.
Since our outer solution only holds for large $f$, we have that the bending energy \linebreak $U_b\ll U_g$. Using the outer solutions (\ref{eqn2:outer_y_cat_sol}) and (\ref{eqn2:outer_y_o_sol}), along with the continuity of tension (\ref{eqn2:outer_sol_b_eqn}), we find 
\begin{align}
     U_\mathrm{total} \approx U_{g} = f \left[ \frac{1}{2} (s^*-1) + \frac{2s^*-1}{4} \log(1-2s^*)\right]. \label{eqn2:inner_sol_energy}
\end{align} 
From this, we see that the branch of solution with $s^* < s^*(a_{\infty})$ has lower energy, and hence is the energetically favourable solution (figure~\ref{fig:energy_outer}(b)).

The outer solution with $s^*\leq s^*(a_{\infty})$ agrees well with the numerical solution far from the point supports (see the plot of the tension in figure~\ref{fig:outer_sol}(b)).
However, it is inconsistent with the continuity of $\theta$ across the point supports, and one can see that the behaviour of the rescaled tension, $t/f$, is dramatically different in the outer solution near the supports than in the  numerics. As mentioned earlier, this is expected because of the small parameter multiplying the highest derivative in (\ref{eqn2:large_f_mom}), which will give rise to a boundary layer-type solution --- we now turn to determine this.

\paragraph{Inner solution: behaviour close to the supports}

To find the inner solution for the behaviour of the PSHE with $f \gg 1$ close to the supports, we will again use equations (\ref{eqn2:large_f_system}) but first rescale the spatial variables near $s=s^*$ and then consider the leading order system. To retain the highest derivative in equation (\ref{eqn2:large_f_mom}) with $f \gg 1$, we scale the spatial variables as follows:
\begin{align}
    s=s^*+f^{-1/2} \zeta, && x= a + f^{-1/2} X, && y=f^{-1/2}Y. \label{eqn:inner_solution_scaling}
\end{align}
Using these scalings, the governing equations (\ref{eqn2:large_f_system}) become, to leading order in $1/f$,
\begin{subequations}
\begin{align}
    \dv{X}{\zeta} &= \cos\theta,  \\
    \dv{Y}{\zeta} &= \sin\theta, \\
    \dv{\zeta}(T \sin\theta + N\cos\theta) &= 0, \label{eqn2:inner_scaled_y}\\
    \dv{\zeta}(T \cos\theta - N\sin\theta) &= 0, \label{eqn2:inner_scaled_x}\\
    \dv[2]{\theta}{\zeta} + N &= 0. \label{eqn2:inner_scaled_theta} 
\end{align} \label{eqn2:inner_sol_scaled_eqns}%
\end{subequations}
To match the inner solution to the outer solution in the limit $f\rightarrow\infty$, and impose the jump condition for $N$ and continuity of $T$ over the supports, we need to consider separate solutions for either side of the support. We will denote each solution by the subscripts $L$ and $R$, for the solution to the left, (catenary) with $\zeta\leq 0$, or right, (hanging chain) with $\zeta \geq 0$, of the support, respectively. We will therefore solve the above equations subject to the following boundary conditions:
\abceqn{eqn2:inner_sol_BCs}{\zeta \rightarrow -\infty: \hspace{13mm} T_L \rightarrow T_c^*, \hspace{13mm} N_L \rightarrow 0, \hspace{13mm} \theta_L \rightarrow \theta_c^*.}
\xyeqn{}{\hspace{9.5mm}\zeta\rightarrow\infty: \hspace{13mm} T_R \rightarrow T_o^*, \hspace{13mm} N_R \rightarrow 0, \hspace{12.75mm} \theta_R \rightarrow \theta_o^*. \eqno{(\theequation{\mathrm{d,e,f}})}}
\xyeqn{}{\hspace{12.5mm}\zeta=0: \hspace{13mm} N_R-N_L = -\sec\theta^*, \hspace{15mm} T_R-T_L=0. \eqno{(\theequation{\mathrm{g,h}})}}
\xyeqn{}{\hspace{12.5mm}\zeta=0: \hspace{13mm} \theta_R-\theta_L=0, \hspace{24mm} \dv{\theta_R}{\zeta}-\dv{\theta_L}{\zeta}=0, \eqno{(\theequation{\mathrm{i,j}})}}
where we have the outer solution tensions and angles at the supports, and use the shorthand $(\cdot)^*=(\cdot)(s^*)$.

Note that $T_c^*=T_o^*$ by continuity (\ref{eqn2:outer_sol_b_eqn}), in the outer solution. 
Solving equations (\ref{eqn2:inner_scaled_y}) and (\ref{eqn2:inner_scaled_x}), subject to the given boundary conditions, we find the solutions for $T$ and $N$ in terms of $\theta^*\leq\theta_L\leq\theta_c^*$ and $-\pi/2\leq\theta_R\leq\theta^*$ to be
\abeqn{eqn2:N_LR_inner_sol_theta}{T_L(\theta_L)=T_c^* \cos\left(\theta_L-\theta_c^*\right), \hspace{16mm} T_R(\theta_R)=-T_o^* \sin\theta_R,}
\xyeqn{}{N_L(\theta_L)=-T_c^* \sin\left(\theta_L-\theta_c^*\right), \hspace{14mm} N_R(\theta_R)=-T_o^* \cos\theta_R. \eqno{(\theequation{\mathrm{c,d}})}}
Using the continuity of $T$, (\ref{eqn2:inner_sol_BCs}h), and jump condition for $N$, (\ref{eqn2:inner_sol_BCs}g), we can then determine the value of $\theta^*$ (matching as in figure~\ref{fig:outer_sol}(a)). To satisfy both conditions, we must have 
\begin{align}
    \tan\theta^* = -\beta. \label{eqn2:frictionless_thetastar}
\end{align}

Using the solutions for $N_{L,R}$, we can solve equation (\ref{eqn2:inner_scaled_theta}) to find $\theta_{L,R}(\zeta_{L,R})$, and hence we have the following solutions for $N_{L,R}(\zeta_{L,R})$ and $T_{L,R}(\zeta_{L,R})$:
\begin{subequations}
\begin{align}
    T_L = T_c^* \left[ 2 \tanh^2 \left(\sqrt{T_c^*} \left( \zeta_L+c_L\right)\right)-1\right], \\
    T_R = T_o^* \left[ 2 \tanh^2 \left(\sqrt{T_o^*} \left( \zeta_R+c_R\right)\right)-1\right], \\
    N_L = -2 T_c^* \tanh \left( \sqrt{T_c^*} \left(\zeta_L+c_L\right)\right) \sech\left( \sqrt{T_c^*} \left(\zeta_L+c_L\right)\right), \\
    N_R = -2 T_o^* \tanh \left( \sqrt{T_o^*} \left(\zeta_R+c_R\right)\right) \sech\left( \sqrt{T_o^*} \left(\zeta_R+c_R\right)\right),
\end{align} \label{eqn2:t_n_inner_sol_xi}%
\end{subequations}
where we have set the constants $c_{L,R}$ such that $\theta_{L,R}=\theta^*$ when $\zeta_{L,R}=0$:
\begin{subequations}
\begin{align}
    c_L=- \frac{1}{\sqrt{T_c^*}} \arctanh\left[\cos \left(\frac{1}{2} (\theta^* - \theta_c^* )\right) \right], \\
    c_R= \frac{1}{\sqrt{T_o^*}} \arctanh \left[\cos\left(\frac{1}{2} \left( \theta^*+\frac{\pi}{2} \right) \right)\right].
\end{align}
\end{subequations}
Note that, since  $\theta^*+\pi/2=-\theta^* +\theta^*_c$, our inner solution has the rotational symmetry, 
\begin{align}
    \theta_R (\zeta_R)-\theta^*=-\theta_L (\zeta_R)+\theta^*,
\end{align}
From this, we have that $c_L=-c_R$. Hence, we also have symmetry in $T_{L,R}$ and $N_{L,R}$.

With the solutions for the inner region, we can now understand the full behaviour of the PSHE in this limiting case. Our inner solution provides a good approximation for the behaviour of the PSHE near the point supports.

\subsection{Regime diagram for the PSHE}

We have seen that the behaviour of the PSHE deforming under increasing $f$ is dependent on the value of the half-gap $a$. Moreover, the value and existence of a maximum force for which a solution can be found, is also dependent on $a$.

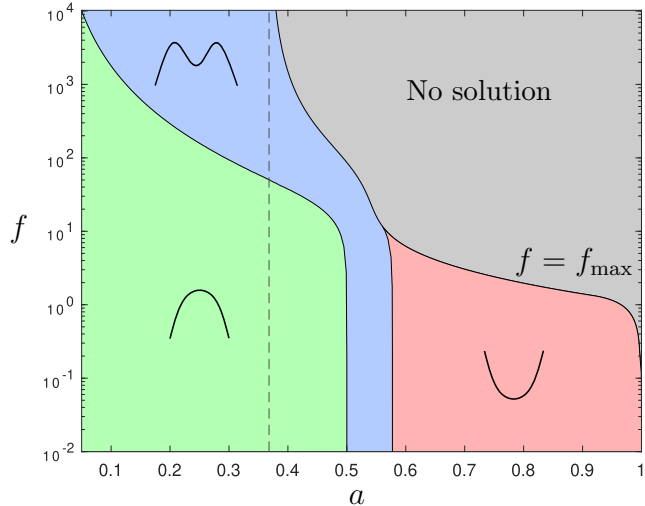
\begin{figure}[ht]
\centering
\begin{overpic}[width=.7\textwidth,tics=10]{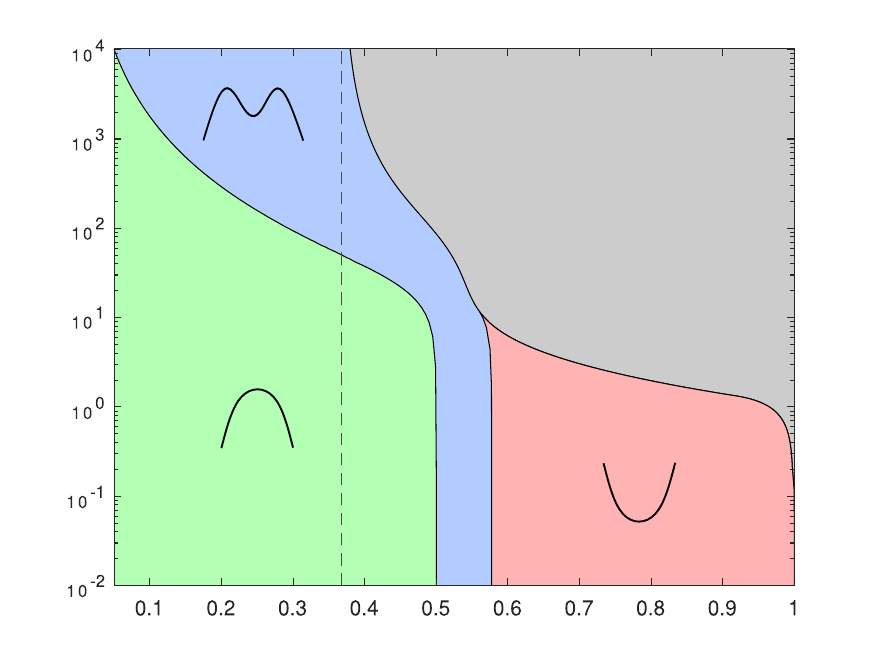}
\put(50,1){$a$ }
\put(3,38){$f$ }
\put(58,57){\small{No solution}}
\put(73,33.5){$f=f_{\mathrm{max}}$}
\end{overpic}
\caption{Phase diagram for increasing force, $f=\hat{f}L^3/(EI)$, against half-gap, $a=\hat{a}/L$. Here the phases are labelled by colour, with the grey region representing no solution, i.e.~the boundary is $f=f_\mathrm{max}$. The asymptote at $a=a_\infty=1/\mathrm{e}$ is given by the dashed line.}\label{fig:phase_diagram}
\end{figure}

We summarize when such a solution can be found for the PSHE, and the shapes of any existing solutions, by considering a phase diagram (figure~\ref{fig:phase_diagram}). As discussed previously, the shape realized by the PSHE is dependent on both $f$ and $a$. In addition to allowing us to see what happens as we change $f$ while holding $a$ fixed, the phase diagram allows us to read off what happens when we change $a$ for a fixed value of $f$, therefore giving us a complete picture of the behaviour of this PSHE problem. 

For the values of $a>1/\mathrm{e}$, the relationships between $f$ and $s^*$ in figure~\ref{fig:f_vs_xi_combined}(a) showed the existence of a finite $f_{\mathrm{max}}$. This maximum corresponds to the applied body force being sufficiently large that the PSHE slips through the supports; when $a \leq 1/\mathrm{e}$, the PSHE can withstand an infinite force without slipping between the supports. Moreover, we find numerically that the value of $f_\mathrm{max}$ increases more rapidly when $a$ decreases into the $m$-shape region (i.e.~$a \lesssim 1/\sqrt{3}$) towards $a=1/\mathrm{e}$, as shown by the boundary curve in figure~\ref{fig:phase_diagram}. 

A PSHE under a sufficiently small fixed force $f$ can realize all three shapes by changing only the half-gap $a$. With this result, this phase diagram displays the robustness of this PSHE problem, since there are several continuations one could run to get to a certain $f$ and $a$ value, but the result remains the same regardless.

\section{The effect of frictional forces}
In the previous section, we investigated how a heavy elastic beam suspended over two point supports behaves under an increasing body force. Our analysis was based on the assumption that the contact between the beam and the supports is frictionless. We found that the critical force required to push the beam completely between the supports (so that no equilibrium exists) is dependent on the size of the gap between the supports: generally, this critical force increases as the gap size reduces. While this result is intuitive, we showed that, below a critical half-gap $a_{\infty}=1/\mathrm{e}$, the PSHE can, in fact, support an arbitrarily large body force.

In reality, however, the PSHE will be affected by frictional forces acting at the contact points. We expect that, with the inclusion of friction in the model, there will be a corresponding increase in the critical load at which the PSHE falls through: the existence of frictional forces should change the required global force balance and allow for more arc length to comfortably be supported between the supports, before falling through. We also anticipate that the inclusion of friction may increase the critical gap width at which the PSHE can withstand an arbitrarily large body force.

\subsection{Model setup}

To incorporate the effects of friction into the model, we use the same governing equations as previously, i.e.~(\ref{eqn2:dimensionless_eqns}). The key difference is that friction allows for a discontinuity in the tension across the supports. We therefore must modify the boundary conditions on the tension and normal forces at the point supports given in (\ref{eqn2:dimensionless_BCs}i,j).

We  impose Coulomb friction and consider a (dimensionless) frictional force $F_{\mu}$, and normal reaction force $R_N$, such that 
\begin{align}
-|\mu|R_N \leq F_{\mu} \leq |\mu|R_N.
\end{align}
Since the frictional force may be in either direction, we take a signed friction coefficient,~$\mu$. 
At the onset of sliding, we will have that $F_{\mu}= \mu R_N$ \citep{Plaut2011supports}.
Since our focus is on the time-independent problem, we focus on when the beam is at the onset of sliding. We will also impose a direction of friction and hold it fixed throughout the evolution, i.e.~the sign of $\mu$ will remain fixed throughout the evolution.

We also note that since our focus is on finding solutions that are on the point of sliding, i.e.~$F_\mu= \mu R_N$, we will actually have a family of solutions with \linebreak $-|\mu| R_N \leq F_\mu \leq + |\mu| R_N$. If $|\mu|=1$, for example, the PSHE could obtain any configuration between those found for $\mu=-1$ and $\mu=1$.

\subsubsection{Governing equations}

In the previous section, we determined the (now dimensionless) reaction forces $R_x$ and $R_y$ by considering a global force balance over half of the PSHE. With friction, the vertical global force balance remains the same as in the frictionless case since we still have that $n(0)=0$ by symmetry; we therefore have $R_y=f$, which is equivalent to the dimensional $\hat{R}_{\hat{y}}$ as before. However, with the contact between the PSHE and the supports now being frictional, we no longer impose that $t$ is continuous. Instead, we have
\abeqn{}{F_{\mu} = -[t]_{s=s^*}, \hspace{35mm}
    R_N = -[n]_{s=s^*}.}
We impose the friction law $F_{\mu}=\mu R_N$, and with $R_y=f$, this gives 
\begin{align}
    R_x=\frac{\tan\theta^*- \mu}{1+\mu\tan\theta^*}f.
\end{align}
The resulting jump conditions are 
\abeqn{eqn3:fric_jump_conditions}{[t]_{s=s^*}=- \frac{f \mu}{\mu\sin\theta^*+ \cos\theta^*}, \hspace{12mm} [n]_{s=s^*}=- \frac{f}{\mu\sin\theta^*+ \cos\theta^*}.}
Note that upon setting $\mu=0$, we recover the jump conditions for the frictionless case (\ref{eqn2:dimensionless_BCs}i,j).

\subsection{Numerical Approach}

As with the frictionless case, we solve the system (\ref{eqn2:dimensionless_eqns}) numerically as discussed in Section~\ref{sec2:numerics}, subject to (\ref{eqn2:dimensionless_BCs}a--h, k--m) and the new jump conditions (\ref{eqn3:fric_jump_conditions}). To do this, we provide a small-deformation solution as our initial guess for the numerical solver. The solution remains the same as in (\ref{eqn2:small_angle_solutions}) up to leading order in $f$, except for $t(x)$, which changes due to the jump condition (\ref{eqn3:fric_jump_conditions}a), giving
\begin{align}
    t(x)=
    \begin{cases}
        f \mu & x<s^*, \\
        0 & x>s^*.
    \end{cases}
\end{align}
We will investigate the effects of such a frictional force by imposing a value of $\mu$ and holding it constant whilst increasing either $f$ or $s^*$ via continuation. Similarly to the frictionless case, we are interested in the critical force needed for the PSHE to fall between the supports, as well as the values of the half-gap below which the PSHE can withstand an arbitrarily large body force without slipping between the supports.

\subsection{Results}
\subsubsection{$f_\mathrm{max}$ and the coefficient of friction}

In the frictionless case, we saw that the size (and even existence) of a finite maximal body force, $f_{\mathrm{max}}$, is dependent on the half-gap $a$ (figure~\ref{fig:f_vs_xi_combined}). For $a \gtrsim 0.5$, $f_{\mathrm{max}}$ is determined by treating $s^*$ as the control parameter and finding the corresponding $f$. However, for $a \lesssim 0.5$, we first increase $f$ via continuation until the PSHE no longer realizes an $n$-shape, and then switch the control parameter to $s^*$.  

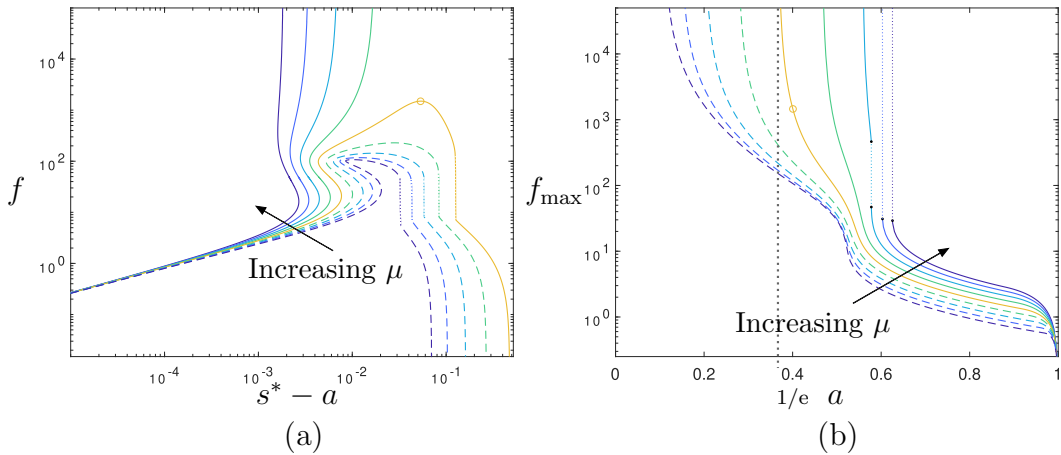
\begin{figure}[ht]
\centering
\begin{overpic}[width=.995\textwidth,tics=10]{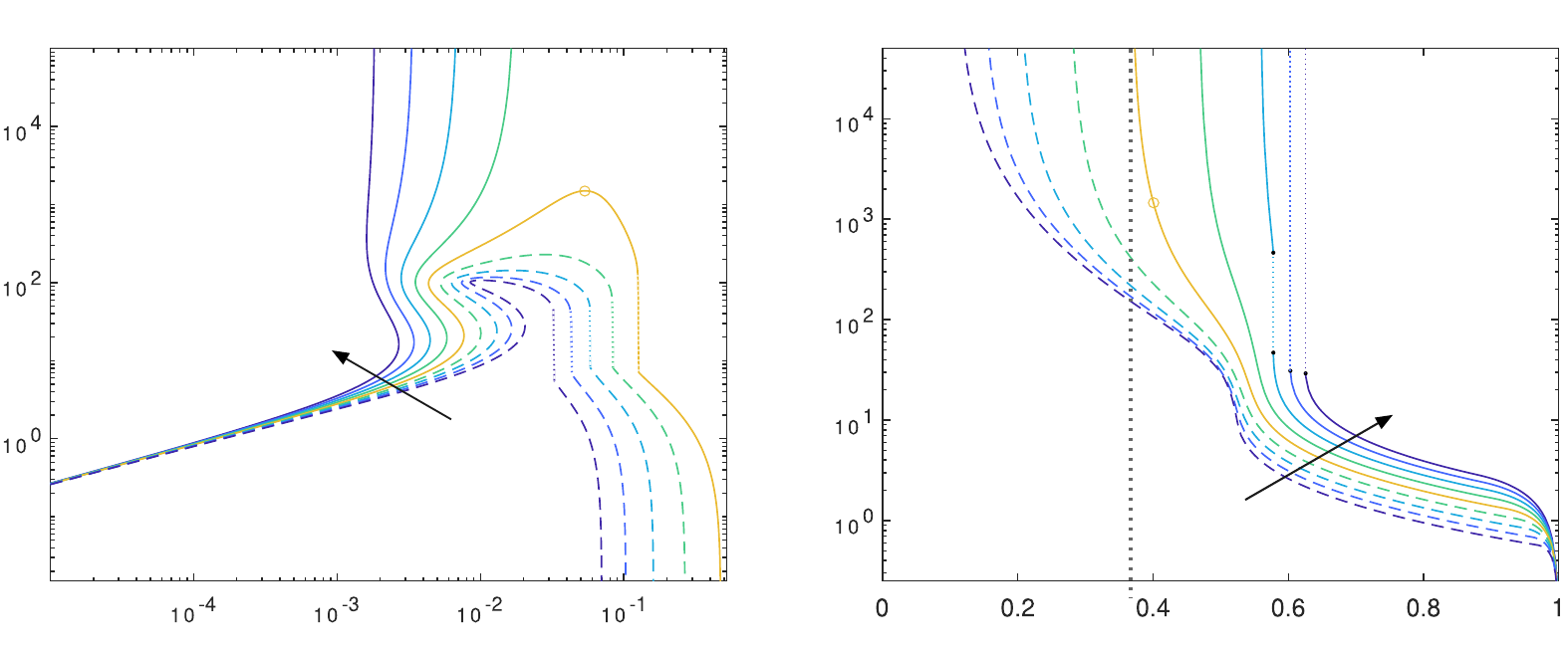}
\put(21.25,0){$s^*-a$}
\put(-3,20){$f$}
\put(20.5,12.25){\small{Increasing $\mu$}}
\put(77,0){$a$}
\put(47.5,20){$f_{\mathrm{max}}$}
\put(68,7){\small{Increasing $\mu$}}
\put(72,0){\scriptsize{$1/\mathrm{e}$}}
\put(24,-4){(a)}
\put(76,-4){(b)}
\end{overpic}
\vspace{1mm}
\caption{(a) Plot of the force, $f$, against scaled excess contact arc length, $s^*-a$, for half-gap $a=0.4$.
(b) Plot of the maximum force, $f_{\mathrm{max}}$, against half-gap, $a$, where the asymptote at $a=1/\mathrm{e}$ is given by the dotted line. In both (a) and (b), results are plotted for $\mu~=~\{-1,-0.75,-0.5,-0.25,0,0.25,0.5,0.75,1\}$, with dashed curves representing $\mu<0$ and solid curves $\mu \geq 0$. The value of $f_\mathrm{max}$ for $a=0.4$ and $\mu=0$, is marked by a circle in (a) and (b).}
\label{fig:fric_max_force}
\end{figure}

With the inclusion of friction, we see that the existence and size of $f_{\mathrm{max}}$ is now also dependent on the value of $\mu$ (figure~\ref{fig:fric_max_force}(a)).
For example, for $a=0.4$ we are not able to find a maximum $f$ for $\mu \geq 0.25$ as chosen in figure~\ref{fig:fric_max_force}(a). For $\mu\leq0$, however, there is a finite $f_{\mathrm{max}}$ which increases with $\mu$. This tells us that, by increasing the frictional force from the supports, the PSHE is able to withstand a greater force. We also see from figure~\ref{fig:fric_max_force}(a) that the value of $s^*_{\mathrm{crit}}=s^*(f_{\mathrm{max}})$ increases with $\mu$, and for a given $f$, the contact arc length $s^*$ decreases with $\mu$.

Figure~\ref{fig:fric_max_force}(b) shows the maximum supportable force $f_{\mathrm{max}}$ (when it exists) as a function of $a$ for different values of $\mu$.  This plot also shows that the value of $a=a_{\infty}(\mu)$ --- the value below which the force may increase without ever causing slip-through --- changes with $\mu$ from the frictionless value of $a_{\infty}(\mu=0)=1/\mathrm{e}$. 
When $\mu>0$, we see that the maximum force appears to blow up for a greater value of half-gap $a_{\infty}$. This supports the expectation that frictional forces would make it more difficult for the PSHE to slip between the supports. Conversely, for $\mu<0$, we see the maximum force blow up for a smaller value of $a_{\infty}$.

For $\mu=0.5$, we find that there is a jump in $f_\mathrm{max}$ from $f_\mathrm{max} \approx 45$ to \linebreak$f_\mathrm{max} \approx 460$ when $a \approx 0.5778$. For any value of $f$ within this range, there exists an equilibrium solution for the PSHE, however we do not see a configuration, for any value of $a$, for which $f_\mathrm{max}$ lies within this range.

We first notice a jump in $f_{\mathrm{max}}$ for $\mu \gtrsim 0.47$ and we see the size of this jump increases with $\mu$, until it becomes seemingly indefinite, i.e.~numerically we can no longer find an $f_\mathrm{max}$ when increasing $f$ up to $10^7$.
For $\mu={0.75,1}$, we find that there is no gradual increase in $f_{\mathrm{max}}$ as $a \rightarrow a_{\infty}$ (figure~\ref{fig:fric_max_force}(b)). Rather, when decreasing $a$, we see an instantaneous change from a finite $f_{\mathrm{max}}$, to the system being able to support an arbitrarily large body force. This is believed to be due to the blow-up values being in a region where the PSHE can only realize a $u$-shape (i.e.~ $a \gtrsim 1/\sqrt{3}$). We would always expect there to exist a finite $f_{\mathrm{max}}$ for a $u$-shape solution, whereas the mechanism for withstanding an infinite body force relies on the overhanging ends being long enough to support the load, i.e.~an $m$-shape. When evolving under increasing $f$, we do not see a natural transition from $u$-shape to $m$-shape for these high $a$ values, which could explain this instantaneous blow-up.

\subsubsection{Behaviour for large $f$ and $a\leq a_\infty$}

As friction only plays a role in the contact region, we can find both outer and inner solutions in the limit of infinite body force for the frictional problem, just as we did in the frictionless case. Since the scaled governing equations remain the same as in (\ref{eqn2:large_f_system}), the form of the solutions in terms of $\beta$, $\theta$, and $\zeta$ will remain the same as in Section~\ref{sec2:large_f}. However, the values of $\beta$ and $\theta^*$ must now  depend on both $a$ and $\mu$ because of the changes in jump conditions for $t$ and $n$ --- compare (\ref{eqn2:dimensionless_BCs}i,j) with (\ref{eqn3:fric_jump_conditions}). 

\paragraph{Outer solution: the frictionally supported catenary}

In the frictionless case, recall that we found $a_{\infty}=1/\mathrm{e}$ by imposing continuity of $T$, i.e.~$T_c^*=T_o^*$, using the outer solutions (\ref{eqn2:t_outer_c}) and (\ref{eqn2:t_outer_o}). However, in the presence of friction, $T$ is no longer continuous, so we must instead impose that $T_o^*-T_c^*$ satisfies the jump condition, (\ref{eqn3:fric_jump_conditions}a), for $t/f$, i.e.~we must now find $\beta$ such that  
\begin{align}
    (1-s^*) - \sqrt{{s^*}^2+\beta^2} =  - \frac{ \mu}{\mu\sin\theta^*+ \cos\theta^*}. \label{eqn3:fric_outer_definition}
\end{align}
Unlike with the frictionless case, we cannot immediately derive from this an equation for $a$, since $\theta^*$ is unknown;  rather, we must first look at the inner solution to derive an equation for $\theta^*$.

\paragraph{Inner solution: behaviour close to the frictional supports}

Whilst the condition that determines $\beta$ has changed, the inner solution is still governed by the same system of equations and boundary conditions as in (\ref{eqn2:inner_sol_scaled_eqns}) and (\ref{eqn2:inner_sol_BCs}). Hence, we will still find the same solutions in terms of $\theta^*$ for $N_{L,R}$ and $T_{L,R}$ (\ref{eqn2:N_LR_inner_sol_theta}), only now  $T_c^* \neq T_o^*$. Using these equations in the jump conditions for $t/f$ and $n/f$, (\ref{eqn3:fric_jump_conditions}), we can then find $\theta^*(\beta;\mu)$. 

Substituting in the solutions (\ref{eqn2:N_LR_inner_sol_theta}), along with $T_c^*$, $T_o^*$, and $\theta_c^*$ from the outer solution, we find that 
\begin{align}
    \tan\theta^* = \frac{\mu-\beta}{\mu \beta +1} = \tan (\phi_f-\arctan\beta), \label{eqn3:inner_thetastar}
\end{align}
i.e.~$\theta^*=\phi_f - \arctan\beta$, where $\phi_f=\arctan\mu$ is the friction angle. We note once again that for $\mu=0$ we return to the frictionless result (\ref{eqn2:frictionless_thetastar}). We also note that, with the imposition of friction, we no longer have rotational symmetry of the solutions due to the change in force balancing.

Further, with equation (\ref{eqn3:inner_thetastar}) for $\tan \theta^*$, we can find an equation for $a(\beta;\mu)$. Substituting (\ref{eqn3:inner_thetastar}) into  (\ref{eqn3:fric_outer_definition}) gives a transcendental equation for $\beta(s^*;\mu)$, \linebreak namely:
\begin{align}
    (1-s^*) - \sqrt{{s^*}^2+\beta^2} = - \mu \sqrt{\frac{\beta^2+1}{\mu^2+1}}.
\end{align}
Eliminating $s^*$ using (\ref{eqn2:arc_length_cat}), we can reduce this to an equation for $a(\beta;\mu)$:
\begin{align}
    a=   \beta \log \Bigg[ \frac{1}{\beta} \left(1+\mu \sqrt{\frac{\beta^2+1}{\mu^2+1}} \right) \Bigg].  \label{eqn3:a_inf_inner_sol}
\end{align}
Note that, again, setting $\mu=0$ returns us to the frictionless result for $a(\beta)$ (\ref{eqn2:a_inner_sol}). As in the frictionless case, the expression $a(\beta;\mu)$ in equation (\ref{eqn3:a_inf_inner_sol}) is multi-valued. 
Hence an $a_{\infty}(\mu)$, the largest value of $a$ for which a catenary solution is possible, exists; to find it, we must solve for the maximum that can be attained by the right-hand side of (\ref{eqn3:a_inf_inner_sol}) as $\beta$ varies with fixed $\mu$. This maximization must be done numerically. For $a<a_\infty$, the same energy argument as in Section~\ref{sec2:outer_sol}, shows that the solution with contact arc length $s^* \leq s^*(a_{\infty})$ is the physically relevant one since the energy is lower than that of the other solution.

\begin{figure}[ht]
\centering
\begin{overpic}[width=.6\textwidth,tics=10]{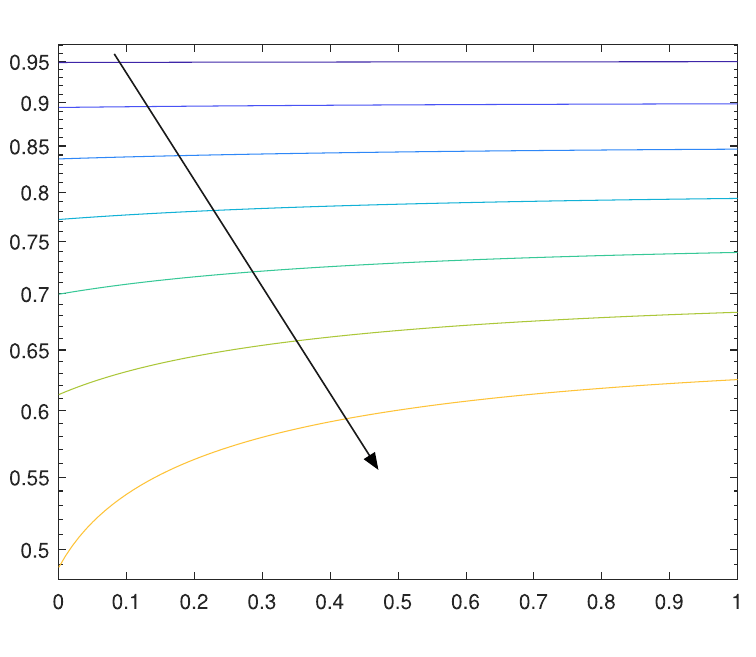}
\put(52,1){$\mu$}
\put(-25,45){$y(0)-y(1)$}
\put(43,20){\small{Increasing $a$}}
\end{overpic}
\caption{Difference in height between the hanging rod's centre and outer edge, $y(0)-y(1)$, as a function of friction coefficient $\mu$, for $a=\{0.05,0.1,0.15,0.2,0.25,0.3,0.35\}$. This height difference is governed by \eqref{eqn3:inner_height} but is affected by the dependence of $\beta$ on $\mu$ and $a$.}
\label{fig:app_change_in_height}
\end{figure}

Finally, we note that $T(0)-T(1)=\beta$, but that now, since $T$ is no longer continuous over the point supports, we can no longer conclude that this is also equal to the difference in height. With the solution for $\theta^*$, (\ref{eqn3:inner_thetastar}), the difference in height now becomes
\begin{align}
     y(0)-y(1) = \beta - \mu \sqrt{\frac{\beta^2+1}{\mu^2+1}}.
      \label{eqn3:inner_height}
\end{align} This difference in height between the centre and edge of the hanging rod is shown for a variety of $a$ and $\mu$ in figure~\ref{fig:app_change_in_height}. This shows that increasing friction \emph{increases} the height difference --- a result that is physically reasonable, but initially appears to be at odds with the `$-$' in the right hand side of \eqref{eqn3:inner_height}, which might suggest the opposite. (The key to this apparent contradiction is that $\beta$ depends also on $\mu$.)

\subsection{Apparent power-law behaviour in $f_\mathrm{max}$}

Whilst the numerical results for $f_{\mathrm{max}}(a;\mu)$ and $a_\infty(\mu)$ give qualitative information, it is natural to ask how $f_\mathrm{max}$ behaves as $a \rightarrow a_\infty$ from above. Given that $f_\mathrm{max}$ must diverge as $a \rightarrow a_\infty$, a natural way to capture this is to investigate power-law relationships:
\begin{align}
    f_{\mathrm{max}}(a;\mu)= \lambda(\mu) \left[ a-a_*(\mu) \right] ^{\gamma(\mu)} ,
\end{align}
for some critical value $a_{\infty}(\mu)$, exponent $\gamma(\mu)$, and constant $\lambda(\mu)$. (We expect $a_*(\mu)=a_{\infty}(\mu)$, but leave this free for the moment.) To determine these parameters, we follow \citet{brun2016pulley}, taking the logarithm of both sides of this equation, and then the derivative with respect to $a$, to find that 
\begin{align}
    \dv{\log f_{\mathrm{max}}(a;\mu) }{a} = \frac{\gamma(\mu)}{a-a_*(\mu)}. \label{eqn3:power_law}
\end{align}
By using the reciprocal of equation (\ref{eqn3:power_law}), we can find best-fit values for both parameters $\gamma$ and $a_*$ using linear regression. 

For example, in the frictionless case, for $a$ near the blow-up value $a_\infty$ we find that the slope is $1/\gamma(0) \approx -0.534$, and note that $8/15 \approx 0.5333$, with a corresponding $x$-intercept of $0.368$. This corresponds to having $\gamma(0) \approx -15/8$ and $a_*(0) \approx 1/\mathrm{e}$. Hence, for $a$ near $a_\infty$, we will approximate $f_{\mathrm{max}}$ by the following power-law:
\begin{align}
    f_\mathrm{max}(a;0)\approx 2.5\left(a-\frac{1}{\mathrm{e}}\right)^{-15/8}. \label{eqn3:frictionless_power}
\end{align}

The same procedure for each value of $\mu$ shows the same linear trend for $(\dv*{\log f_\mathrm{max}(a;\mu)}{a})^{-1}$ as $a$ tends to some $a_{\infty}(\mu)$. This suggests that $f_{\mathrm{max}}(a;\mu)$ follows the same power-law exponent as in (\ref{eqn3:frictionless_power}), for each value of $\mu$, i.e.~
\begin{align}
    f_{\mathrm{max}}(a;\mu)= \lambda(\mu) \left[a-a_*(\mu)\right]^{-15/8}. \label{eqn3:power_law_fric}
\end{align}
We emphasize that the expression in (\ref{eqn3:frictionless_power}) and (\ref{eqn3:power_law_fric}) is based purely on numerical observation; we do not understand the origin of the rational exponent $15/8$ but it is numerically robust. 

\begin{figure}[ht]
\centering
\begin{overpic}[width=.995\textwidth,tics=10]{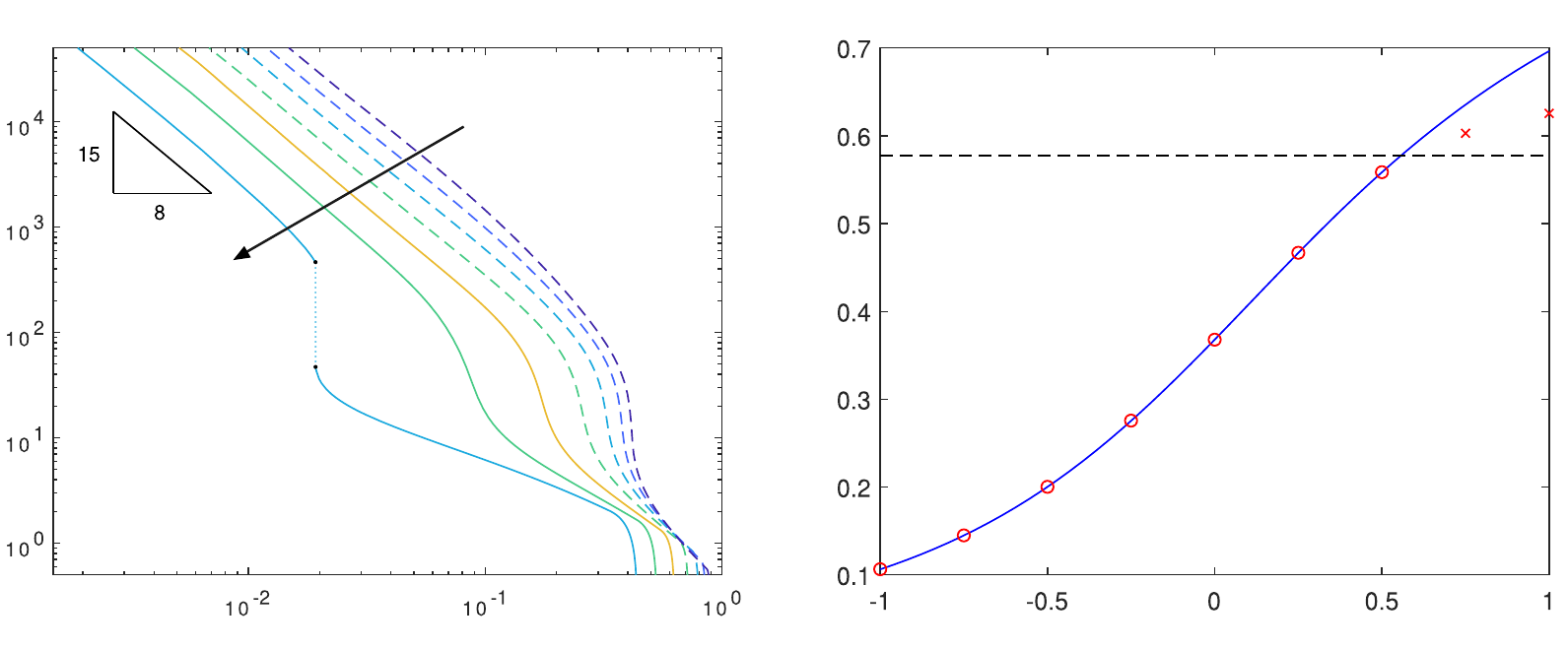}
\put(-6,20){\small{$f_{\mathrm{max}}$}}
\put(20,-1){\small{$a-a_*(\mu)$}}
\put(28,34.5){\small{Increasing $\mu$}}
\put(77,-1){\small{$\mu$}}
\put(49,20){\small{$a_*$}}
\put(58,33.5){\footnotesize{$a = \frac{1}{\sqrt{3}}$}}
\put(24,-5){{(a)}}
\put(76,-5){{(b)}}
\end{overpic}
\vspace{2mm}
\caption{(a) Plot of the maximum force, $f_{\mathrm{max}}(a;\mu)$, against scaled half-gap, $a-a_*(\mu)$, for $\mu~=~\{-1,-0.75,-0.5,-0.25,0,0.25,0.5\}$, where the dashed curves represent results with $\mu<0$ while solid curves represent $\mu \geq 0$. (b)  Plot of the critical half-gap, $a_*(\mu)$, for which a catenary-type solution may be found as $f \rightarrow \infty$. The prediction from the inner solution incorporating friction is given by the solid curve. The values of $a_*$ obtained from the power-law approximation are given for $\mu~=~\{-1,-0.75,-0.5,-0.25,0,0.25,0.5\}$ by the circles. The estimated values from the numerics found for $\mu~=~\{0.75,1\}$ are given by the crosses, but are believed to deviate from the analytical prediction because the corresponding values for $a_*$ lie in the region where only a $u$-shape can be realized (i.e.~$a \gtrsim 1/\sqrt{3}$), as indicated by the dashed line.}
\label{fig:power_and_a_infty}
\end{figure}

Under the assumption that $f_{\mathrm{max}}(a;\mu)$ obeys the same power-law for each value of $\mu$, we can fine tune our estimates for $a_*(\mu)$ such that this holds true.  We can see from figure~\ref{fig:power_and_a_infty}(a) that, for $\mu \leq 0.5$, with a good numerical guess for $a_*$, the maximum force indeed seems to be following the same power-law.

The values of $a_*$ calculated this way compare well with the values of $a$ at which $f_{\mathrm{max}}(a;\mu) \rightarrow \infty$, i.e.~$a_{\infty}(\mu)$, as estimated from the asymptotics (\ref{eqn3:a_inf_inner_sol}) for $-1\leq\mu\leq0.5$ (figure~\ref{fig:power_and_a_infty}(b)). However, for $\mu \gtrsim 0.5$, this agreement breaks down, as $a_{\infty}$ is above the flip-point for these large values of $\mu$. The inner and outer solutions we have calculated do not take into account that the PSHE will be evolving from a $u$-shape under increasing force. 
The two values of $a_{\infty}$ for $\mu=0.75$ and $\mu=1$ were instead estimated by running continuations in $f$, decreasing $a$ until the continuation could run for an arbitrarily large $f$.

\section{Experiments}

We performed model experiments to test two key predictions of our modelling --- the transitions between different shapes as $f$ and $a$ vary and the presence of a  critical $a$ below which very large $f$ can be supported (though of course it is impossible to reach arbitrarily large $f$ experimentally).

\subsection{System setup and rod fabrication}

The key experimental parameter is the dimensionless body force, $f$. Using our non-dimensionalization of  $\hat{f}$ \eqref{eqn2:non-dim scaling}, the dimensionless gravitational force on a cylindrical rod of length $L$ and radius $r$ is
\begin{align}
    f= \frac{4 \rho g L^3}{E r^2},
    \label{eqn:dimlessF}
\end{align}
where $\rho$ is the volumetric density, $g$ is the acceleration due to gravity, $E$ is the Young's modulus, and $I=\pi r^4/4$ is the second moment of inertia of the rod's cross-section. 

The expression in \eqref{eqn:dimlessF} shows that reaching large values of $f$ requires the fabrication of slender rods ($L/r\gg1$) with very low Young's modulus but appreciable density. To do this, we fabricated cylindrical rods of Polyvinyl Siloxane (PVS, Elite Double 32 from Zhermack) by injecting the mixed, liquid form into cylindrical tubes using a syringe pump and allowing the rod to cure. The rod could then be removed by carefully pulling from one end.

The density and Young's modulus of Elite Double 32 were measured to be $\rho = 1150 \,\si{\kg.\m^{-3}}$ and $E=8\times10^5 \,\si{\pascal}$, respectively, agreeing with previously reported values \cite[see][for example]{Box2020}. (While other types of Elite Double have lower values of $E$, the resulting solids were found to extend noticeably under their own weight, in contrast to our inextensible elastica model. Similarly, they were very `tacky',  making it difficult to remove from the cylindrical mould.)

The value of $f$ was varied by changing the dimensions of the rod used experimentally; specifically two different cylinder radii, $r=1.5\,\si{\mm}$ and $r=4.5\,\si{\mm}$ were used with $L$ chosen to obtain the desired $f$. To span a large range of $f$, we used lengths $L$ such that  $\log_{10}f={0,1,2,3}$, as well as the largest attainable $f \approx 24300$.

\begin{figure}[ht]
\centering
\begin{overpic}[width=.9\textwidth,tics=10]{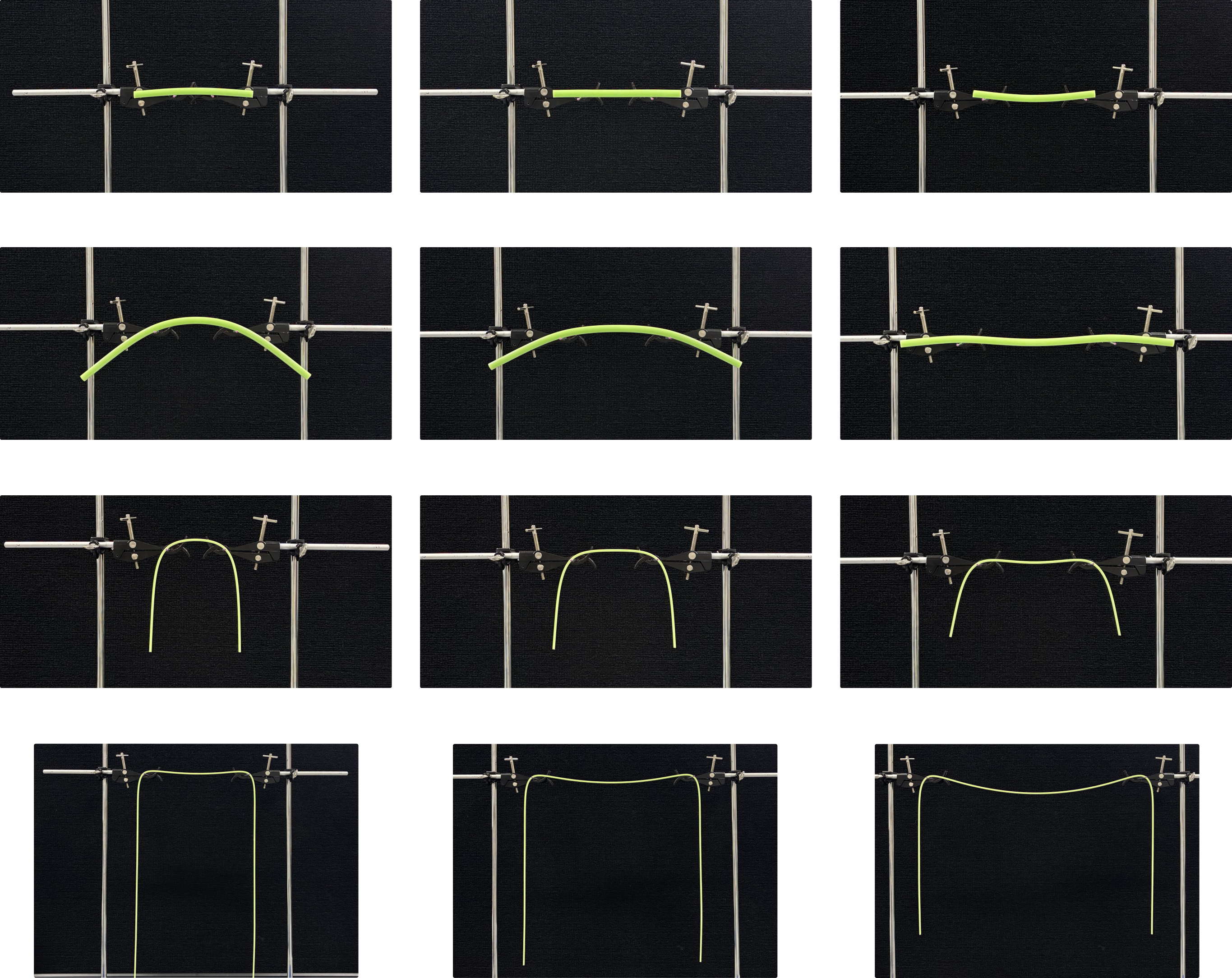}
\put(46,20){\scriptsize{$a=0.3$}}
\put(46,40){\scriptsize{$a=0.3$}}
\put(46,60){\scriptsize{$a=0.4$}}
\put(45.5,80){\scriptsize{$a=0.55$}}
\put(12.25,20){\scriptsize{$a=0.2$}}
\put(12.25,40){\scriptsize{$a=0.2$}}
\put(12.25,60){\scriptsize{$a=0.3$}}
\put(12.25,80){\scriptsize{$a=0.3$}}
\put(80.25,20){\scriptsize{$a=0.4$}}
\put(80.25,40){\scriptsize{$a=0.4$}}
\put(80,60){\scriptsize{$a=0.55$}}
\put(80.25,80){\scriptsize{$a=0.8$}}
\put(-5,71){\scriptsize{(a)}}
\put(-5,51){\scriptsize{(b)}}
\put(-5,31){\scriptsize{(c)}}
\put(-5,10){\scriptsize{(d)}}
\end{overpic}
\caption{Typical  experimental images showing a range of different deformed rod shapes as the body force and support-separation, $f$ and $a$, are varied. In each row, the value of $f$ is fixed: (a) $f\approx 10$, (b) $f\approx100$, (c) $f\approx1000$, (d) $f\approx24300$. (Note that to vary $f$, the rod diameter and length are varied. For scale, the diameter of the vertical stand in each image is $1.05\,\si{\cm}$.)  }\label{fig:exp_photos}
\end{figure}
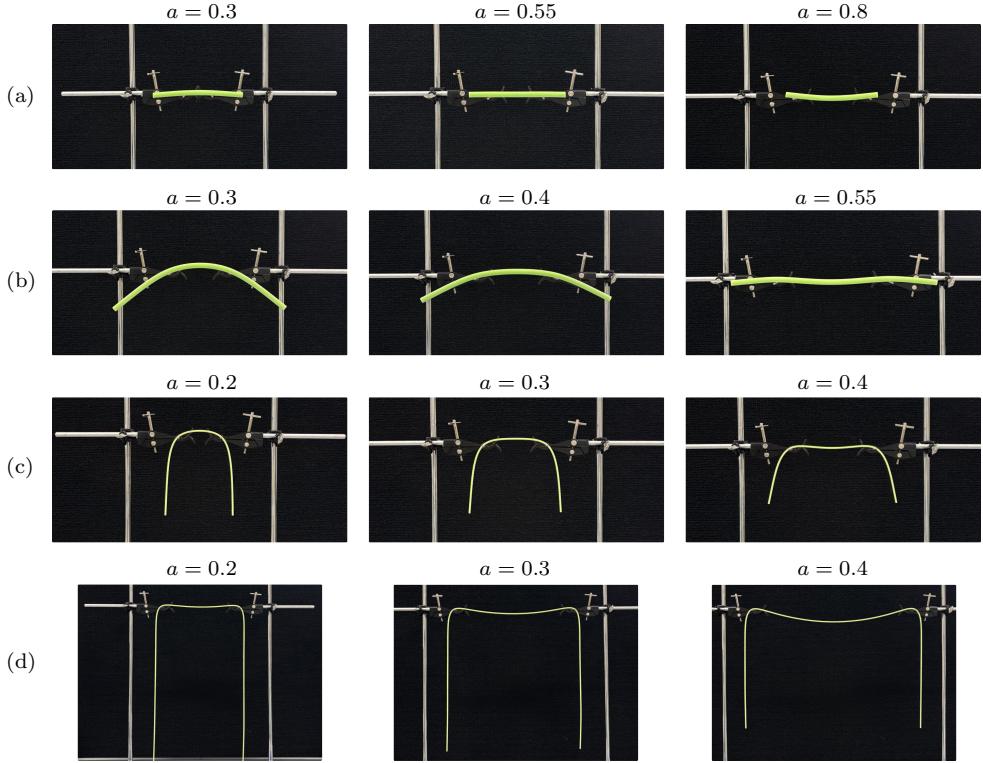

Point-like supports were made by clamping  glass capillary tubes  ($\diameter= 1\si{\,\mm}$) horizontally (see images in figure \ref{fig:exp_photos}). To reduce the friction between the rod and the supports, we apply oil (Johnson's baby oil) to the supports; experiments on planar PVS samples on larger glass surfaces suggested a sliding angle $\phi\in(5,8)^\circ$, corresponding to a friction coefficient $\mu \in (0.08,0.15)$.  The distance between the two supports, rescaled by the rod's length, gives the corresponding value of $a$. 
 
 The rod is placed by hand onto the supports as symmetrically as possible (as judged by eye) and held in place until the effects of inertia have dispersed; in this way, the experimentalist can `feel' whether the rod is close to equilibrium before releasing it. (Releasing without this leads to large perturbations that can destabilize scenarios for which equilibrium is actually possible.) If the rod remains supported and in equilibrium for a given $a$ then the shape of deformation is noted\footnote{However, for small $f\approx1$ it was very difficult to distinguish the assumed shapes from the `u'-shape.} (and an image captured from the side view). The gap-width $a$ is then increased by $0.1$ until slip-through is reached.  We then introduce smaller changes to $a$ to determine the  largest value of $a$ (with fixed $f$) for which an equilibrium state exists to two significant figures; we denote this quantity by  $a_{\mathrm{max}}(f)$.

 \subsection{Results}

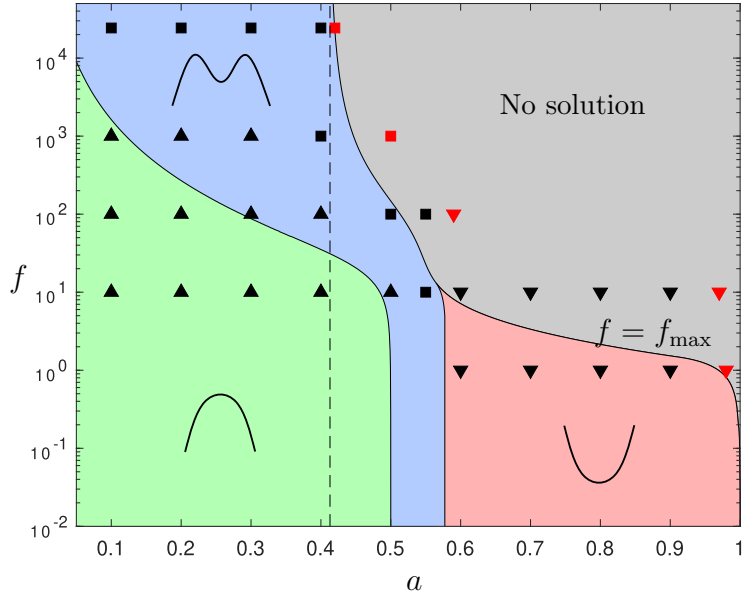
\begin{figure}[ht]
\centering
\begin{overpic}[width=.7\textwidth,tics=10]{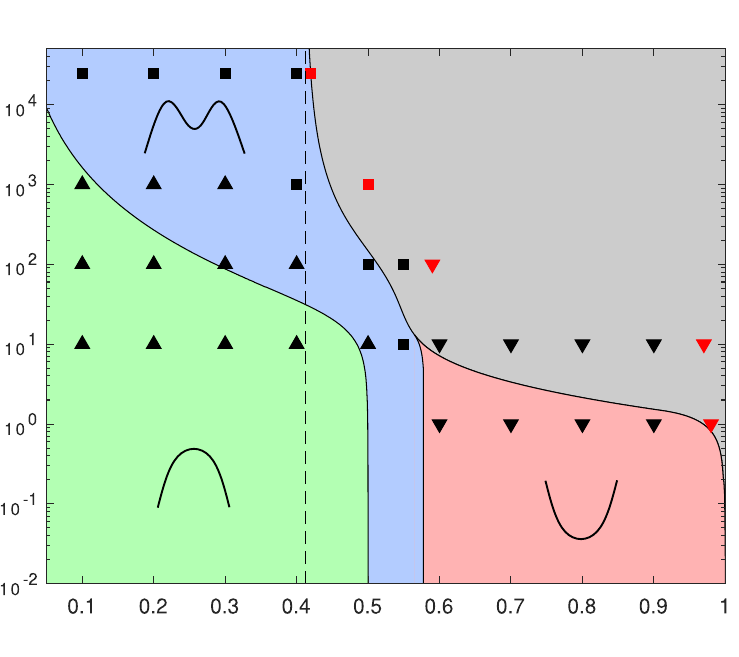}
\put(52,1){$a$}
\put(-3,43){$f$}
\put(65,67){\small{No solution}}
\put(78,35.5){$f=f_{\mathrm{max}}$}
\end{overpic}
\caption{Phase diagram for body force, $f$, against half-gap, $a$, for $\mu=0.115$. Here the phases are labelled by colour, with the grey region representing no solution, i.e.~the boundary is $f=f_\mathrm{max}$. The asymptote at $a=a_\infty(\mu=0.115)=0.413$ is given by the dashed line. The different shapes are labelled by $\blacktriangle:n$-shape, $\blacksquare:m$-shape, $\blacktriangledown:u$-shape. For each $f$, the largest equilibrium $a$, $a=a_{\mathrm{max}}(f)$, is marked  in red.}
\label{fig:phase_diagram_exp}
\end{figure}

Typical experimental images are shown in figure~\ref{fig:exp_photos}. These demonstrate examples of all three of the different deformed shapes as $f$ and $a$ are varied. The particular shape adopted for each pair, $(a,f)$ is also noted in a regime diagram (figure~\ref{fig:phase_diagram_exp}). This is compared with the theoretically predicted regime diagram for $\mu=0.115$. (This value of $\mu$ was obtained by comparing  the maximum value of $a$ with the $a_{\mathrm{max}}$ found from experiments for $f=24300$. Note that this value of $\mu$ lies within the experimentally-determined range of $\mu$ values.)

The phase diagram in figure \ref{fig:phase_diagram_exp} shows qualitatively similar behaviour between experiments and the theoretical predictions. In particular, shape changes occur as $f$ and/or $a$ are changed. Moreover, the prediction that arbitrarily large $f$ may be supported as $a$ decreases below $a\approx0.4$ seems consistent with our experimental data.  However, the quantitative values of parameters at which these transitions occur differ somewhat. (The most dramatic example of this seems to be the behaviour at $f=10$ where we observe `u'-shapes over a significantly larger range of $a$-values than is expected.) We believe these discrepancies to be a result of the model's assumption that the rod is slender ($r/L\ll1$), which is violated for the thicker, shorter rods required to reach $f\sim10$. Nevertheless, we still see the dramatic decrease in $a_{\mathrm{max}}$ from $a=0.97$ to $a=0.59$ as $f$ increases from $10$ to $100$. We also find for extremely large $f$ that only $m$-shapes are observed and the the experimentally observed value $a_{\mathrm{max}}\approx0.42$, is close to $a=1/\mathrm{e}$ as expected for this small $\mu$ experiment.

\section{Conclusions}

In this paper, we have investigated the problem of a point-supported heavy elastica (PSHE) deforming under the action of a uniform, increasing body force. Our study was inspired by the industrial process of the filtration of fibres and so we sought to understand the behaviour of such an elastic object, focusing on the critical force at which equilibria cease to exist. We then included the effects of friction into the model to investigate how these results change with frictional force.
    
    We showed that the PSHE can adopt three, qualitatively different, shapes during its deformation: depending on the size of the half-gap and the load, $u$, $n$, and $m$-shapes are observed, as summarized in figure~\ref{fig:phase_diagram}. This is supported by the results from our experiments (figures~\ref{fig:exp_photos} and \ref{fig:phase_diagram_exp}). For a fixed half-gap $a$, we found the maximum force $f_{\mathrm{max}}$ that could be withstood by the PSHE, before it falls between the supports, both in the absence and presence of friction (figure~\ref{fig:fric_max_force}(b)). The value of $f_{\mathrm{max}}$ for a given half-gap increases with the inclusion of a frictional force, supporting the expectation that the presence of friction allows for a heavier beam to be supported (figure~\ref{fig:fric_max_force}(a)). The value of $f_{\mathrm{max}}$ also increases as $a$ decreases, implying that the smaller the separation of the supports, the heavier the beam that can be supported.
In the context of a fibre in a filter, this supports the intuitive notion that the finer the filter mesh, the greater the force that can be withstood by the fibre before slipping through.  
Surprisingly, however, we found that there is a critical value of half-gap, denoted $a_{\infty}(\mu)$, below which the PSHE can support an arbitrarily large body force. Via asymptotic analysis, we were able to determine $a_{\infty}$ for a given value of $\mu$ (figure~\ref{fig:power_and_a_infty}(b)). 
We also found numerically that the maximum force $f_{\mathrm{max}}(a;\mu)$ appears to follow the same power-law for each value of $\mu$, with $f_{\mathrm{max}} \sim \left[a-a_{\infty} \right]^{-15/8}$ as $a \searrow a_{\infty}$ (figure~\ref{fig:power_and_a_infty}(a)).

The results from our PSHE provide a good foundation for further work in modelling the microscale filtration of a fibre. This is significant for filtration as our model suggests that to ensure no fibre of length $L>L_\mathrm{min}$ escapes the filter, a pore of size $a_\mathrm{pore}<L_\mathrm{min}/\mathrm{e}$ should be chosen.
Of course, our analysis assumed that the system remains left-right symmetric --- one next step would be relaxing this assumption of symmetry.
In reality, the chances of a fibre lying on a pore of a filter with perfect symmetry is low, so one extension to consider is asymmetric contact, to see how the critical half-gaps and forces change if the PSHE is placed off-centre over the supports. Similarly, our model of friction has assumed that contact with supports occurs only at isolated points. An interesting direction for future research would be to account for the finite radius of curvature of the supports, which would allow for the magnitude and direction of the friction force to vary spatially.

\vspace{5mm}
\noindent\textbf{Acknowledgments}\\
We are grateful for the support of a Christchurch Graduate Studentship (GKC). For the purpose of Open Access, the authors will apply a CC BY public copyright license to any Author
Accepted Manuscript version arising from this submission.

\renewcommand{\bibname}{References}

\end{document}